\documentclass[10pt, a4paper]{article}				
\usepackage[plainpages=true, hypertexnames=true, hyperindex, breaklinks=true, pdfborder={0 0 0}]{hyperref}
\usepackage{bookmark}
\usepackage[utf8x]{inputenc}
\usepackage{ucs}
\usepackage{color}
\usepackage{amsmath}
\usepackage{amsfonts}
\usepackage{amssymb}
\usepackage[ngerman, english]{babel}
\usepackage{mathtools}
\usepackage{times}
\usepackage{colonequals}
\usepackage{makeidx}

\usepackage[left=2cm,right=2cm,top=2cm,bottom=2cm]{geometry} 
\usepackage{fancyhdr}										
\newcommand{\dif}{\;\mathrm{d}}
\newcommand{\Span}{\mathrm{Span}}
\newcommand{\bra}{\langle}
\newcommand{\ket}{\rangle}
\newcommand{\R}{\mathbb{R}}
\newcommand{\C}{\mathbb{C}}
\newcommand{\Ham}{\mathbb{H}}
\newcommand{\J}{\mathcal{J}}
\title{Hypercomplex Algebras and their application to the mathematical formulation of Quantum Theory}

\author{
Torsten Hertig{\small $~^{\mathrm{I}1}$}, Philip Höhmann{\small $~^{\mathrm{II}2}$}, Ralf Otte{\small $~^{\mathrm{I}3}$}
\vspace{1.6mm}\\
\fontsize{10}{10}\selectfont\rmfamily\itshape
$~^\mathrm{I}$\,tecData AG\\
\fontsize{10}{10}\selectfont\rmfamily\itshape
Bahnhofsstrasse 114, CH-9240 Uzwil, Schweiz\\
\fontsize{9}{9}\selectfont\ttfamily\upshape
$~^1$\,torsten.hertig@tecdata.ch\\
\fontsize{9}{9}\selectfont\ttfamily\upshape
$~^3$\,ralf.otte@buhlergroup.com
\vspace{1.2mm}\\
\fontsize{10}{10}\selectfont\rmfamily\itshape
$~^\mathrm{II}$\,info-key GmbH \& Co. KG\\
\fontsize{10}{10}\selectfont\rmfamily\itshape
Heinz-Fangman-Straße 2, DE-42287 Wuppertal, Deutschland\\
\fontsize{9}{9}\selectfont\ttfamily\upshape
$~^2$\,hoehmann@info-key.de
}
\date{March 31, 2014}

\begin{document}
\begin{titlepage}
\maketitle

\thispagestyle{empty}

\begin{abstract}
\noindent
Quantum theory (QT) which is one of the basic theories of physics, namely in terms of \textsc{Erwin Schrödinger}'s 1926 wave functions in general requires the field $\mathbb{C}$ of the complex numbers to be formulated.\\
However, even the complex-valued description soon turned out to be insufficient. Incorporating \textsc{Einstein}'s theory of Special Relativity (SR) (\textsc{Schrödinger}, \textsc{Oskar Klein}, \textsc{Walter Gordon}, 1926, \textsc{Paul Dirac} 1928) leads to an equation which requires some coefficients which can neither be real nor complex but rather must be hypercomplex. It is conventional to write down the \textsc{Dirac} equation using pairwise anti-commuting matrices. However, a unitary ring of square matrices \emph{is} a hypercomplex algebra by definition, namely an associative one.
However, it is the algebraic properties of the elements and their relations to one another, rather than their precise form as matrices which is important. This encourages us to replace the matrix formulation by a more symbolic one of the single elements as linear combinations of some basis elements. In the case of the \textsc{Dirac} equation, these elements are called biquaternions, also known as quaternions over the complex numbers.\\
As an algebra over $\mathbb{R}$, the biquaternions are eight-dimensional; as subalgebras, this algebra contains the division ring $\mathbb{H}$ of the quaternions at one hand and the algebra $\mathbb{C}\otimes\mathbb{C}$ of the bicomplex numbers at the other, the latter being commutative in contrast to $\mathbb{H}$. As it will later turn out, $\mathbb{C}\otimes\mathbb{C}$ contains several \emph{pure non-real} subalgebras which are isomorphic to $\mathbb{C}$, letting bicomplex-valued wave functions be considered as composed from facultatively independent quasi-complex-valued wave functions.
\\
Within this paper, we first consider briefly the basics of the non-relativistic and the relativistic quantum theory.
Then we introduce general hypercomplex algebras and also show how a relativistic quantum equation like \textsc{Dirac}'s one can be formulated using hypercomplex coefficients.
Subsequently, some algebraic preconditions for some operations within hypercomplex algebras and their subalgebras will be examined. For our purpose, an exponential function should be able to express oscillations, and equations akin the \textsc{Schrödinger}'s one should be able to be set up and solved. 
Further, like within $\C$, functions of complementary variables (such like position and momentum) should be \textsc{Fourier} transforms of each other. All this should also be possible within a purely non-real subspace. It will turn out that such a subspace also must be a sub\emph{algebra}, i.e. it must be closed under multiplication. Furthermore, it is an ideal and hence denoted by $\J$. It must be isomorphic to $\C$, hence containing an \emph{internal identity element}. The bicomplex numbers will turn out to fulfil these preconditions, and therefore, the formalism of QT can be developed within its subalgebras.
We also show that the bicomplex numbers encourage the definition of several different kinds of conjugates. One of these treats the elements of $\J$ precisely as the usual complex conjugate treats complex numbers. This defines a quantity what we call a modulus which, in contrast to the complex absolute square, remains non-real (but can be called `pseudo-real'). However, we do not conduct an explicit physical interpretation here but we leave this to future examinations.
\end{abstract}

\paragraph*{keywords} algebras, bicomplex, hypercomplex, quantum mechanics, quantum theory, quaternions, \textsc{Schrödinger} equation, special relativity, wave functions.

\end{titlepage}

\selectlanguage{english}
\section{Introduction}
The history of quantum theory starts with the discovery of the wave-particle-dualism of light (in the broadest sense) by \textsc{Max Planck} (explanation of black-body-radiation, 1900) and \textsc{Albert Einstein} (explanation of the photoelectric effect, 1905). It means that electromagnetic radiation of frequency $\nu$ respectively the pulsatance (angular frequency) $\omega=2\pi\nu$ can be absorbed or emitted in `portions' or quanta of
$
E=h\nu=\hbar\omega
$
only, where $h$ is \textsc{Planck}'s constant (or quantum of action) and $\hbar=\frac{h}{2\pi}\approx 1,054\times 10^{-34}\mathrm{Nms}$ is called reduced \textsc{Planck}'s constant or \emph{Dirac}'s constant. This dualism, however, is not confined to electromagnetic radiation: Searching for a plausible explanation for the stability of election states within an atom, \textsc{Louis Victor de Broglie} applied this dualism to matter in 1924, postulating that to any particle of energy $E$ and momentum $\vec{p}$, a pulsatance
$
\omega=\frac{E}{\hbar}
$
and the wave vector
$
\vec{k}=\frac{\vec{p}}{\hbar}
$
can be attributed.
\paragraph{The wave equation and its complex ansatz for a solution}
\textsc{Erwin Schrödinger} seized \textsc{de Broglies} idea in 1926. He replaced the classical variables by differential operators to develop one of the most important basic equations of quantum mechanics (QM), the wave functions $\phi(\vec{x},t)$ being its solutions. The general real solution
\begin{equation}
a\cos(\vec{k}\cdot\vec{x}-\omega t)+b\sin(\vec{k}\cdot\vec{x}-\omega t),\qquad a,b\in\mathbb{R}
\end{equation}
turned out as unable to solve the equation not least because it is of 1st order in time derivative which requires some kind of exponential function to solve it. Due to \textsc{Leonhard Euler}'s formula
$
e^{i\varphi}=\cos(\varphi)+i\cdot\sin(\varphi)
$, the complex-valued ansatz
\begin{equation}\label{ansatz1}
ze^{i(\vec{k}\cdot\vec{x}-\omega t)}=ze^{\frac{i}{\hbar}(\vec{p}\cdot\vec{x}-Et))},\qquad z\in\mathbb{C}.
\end{equation}
turns out to be apt because it unifies the trigonometric functions with exponential functions and thus solves linear differential equations of different orders including the 1st and the 2nd.\footnote{As relativistic QT shows, a real equation with special real $2\times 2$-matrix coefficients and a real wave function 2 component vector as a solution would work as well, though less elegant. However, the coefficients then were \emph{isomorphic} to complex numbers.}

\paragraph{Interpretation of the wave function}
Not least because of their complex (hypercomplex, respectively) values, a lively debate on the nature of these wave functions soon arose. \textsc{Schrödinger} considered them as representations of physical waves at that time, he thought e.g. of a distribution of charge density.\\
However, the majority of physicists disagreed. Within the year, \textsc{Max Born} suggested an interpretation for the absolute square of the wave function as the probability density which is still valid today. This lead to the Kopenhagen interpretation which in some aspects seem akin to positivism. Its most famous proponent \textsc{Niels Bohr} regarded the wave function as nothing but a useful mathematical aid without any physical reality.\\
In this point, we disagree. We consider complex-valued functions to have special physical properties and being much more than just a mathematical aid \cite{Otte}, for we are convinced that no such aid or pure formalism could have real physical effects, e.g. in form of destructive interference requiring the wave functions themselves to interfere, not simply the probabilities. This \emph{quantum realism} also holds for hypercomplex approaches, these obviously being inevitable for  a consistent depiction of nature. Note that it is false to identify \textit{real values} with \textit{measurable} and \textit{imaginary values} with \textit{not measurable}; the real part of a wave function is as little measurable as its imaginary part. Actually, the only thing to measure are eigenvalues of Hermitian operators; however, it is possible to reconstruct probability densities, i.e. the absolute squares discussed above by multiple measurements on identically prepared quantum systems. It is the \emph{phase} which remains unknown.\footnote{An exception may be some special states in photons known as coherent states; in this case, the number of `particles' is not sharply defined.}

\paragraph{Special relativity and hypercomplex extensions}
Roughly at the same time as QT, the special relativity theory (SRT, \textsc{Einstein}, 1905, see appendix \ref{SRT}) came to existence as an offspring of the cognizance that like to the laws of mechanics, \textsc{Galilei}'s principle of relativity also applies to \textsc{James Clerk Maxwell}'s electrodynamics which implies that  $c=299792458\frac{\mathrm{m}}{\mathrm{s}}$\footnote{this is the today value which is exact by definition since the redefinition of the meter by GCPM in 1983, being within the last error (1973).}, the vacuum speed of light and other electromagnetic waves has to be the same in any inertial system, being independent of its velocity.
Involving SRT in wave mechanics, the scalar complex ansatz turned out to be insufficient for the purpose of fully describing matter. The problem was solved by \textsc{Paul Dirac} in 1928 by setting up an equation with \emph{hypercomplex} coeffizients. These are written as quadratic matrices, while the equation's solution are vectors of functions.

\paragraph{Conventions for the following text}
Universal constants like $c$ or $\hbar$ are actually artifacts of the measuring system (see appendix \ref{SRT}) and don't reveal anything deeper about mathematical relations. Therefore theoretical physicists prefer \emph{natural units} in which they are equal to unity or at least a simple dimensionless number. So we do, using a system of measurement with $\hbar=1, c=1$ unless an exception is explicitly indicated. So, \eqref{ansatz1} becomes
\begin{equation}\label{ansatz2}
ze^{i(\vec{p}\cdot\vec{x}-Et)}.
\end{equation}
In conformity with the conventions of relativity theory, especially general relativity, we further use Greek indices if the set of indices includes zero and Latin ones otherwise. Double indices, especially when one of them is an upper (not to be confused with powers!) and one is a lower, will be summed over unless explicitly negated.
Integrals without bounds are not to be taken as \emph{indefinite} but as \emph{improper}, i.e. the integration is to be calculated over the entire range of the integrand.
Last, we write operators of the form
\begin{equation*}
\frac{\partial}{\partial x}, \frac{\partial^2}{\partial x^2},\frac{\partial}{\partial t}, \frac{\partial^2}{\partial t^2},\cdots
\end{equation*}
in a space-saving manner like  $\partial_x,\partial_x^2,\partial_t,\partial_t^2,\dots$ unless there is anyway a fraction.

\section{Matrix mechanics and wave mechanics}
QT is formulated in two manners which look profoundly dissimilar at the first sight: matrix mechanics (\textsc{Werner Heisenberg}  et al., 1925) and wave mechanics (\textsc{Erwin Schrödinger} et al., 1926). \textsc{Schrödinger}, indeed, proved both manners as equivalent \cite{schr19260318,schr19260127,schr19260223}.
\newline
Matrix mechanics is more general and \emph{coordinate-independent}. It deserves primacy in respect of that any wave mechanics have to be expressible in terms of matrix mechanics\footnote{and indeed is whereas the opposite is not always possible; e.g. there is no position representation of a spin state}, and it provides all concepts and formalism described in appendix \ref{QTideas}. In \ref{anmQT}, a two-state-system and a space of wave functions (in position representation) are shown as two mostly different examples of  \textsc{Hilbert} spaces, i.e. spaces of quantum states.  
Wave mechanics is hence a special case of matrix mechanics. However, it is more graphic since it describes a ``particle'' by functions in space and time. Additionally, it promotes the usage of complex-valued functions which correspons to our purpose of a hypercomplex extension of QT; this is why we mainly consider it below.

\subsection{The \textsc{Schrödinger} equation and its solutions}
\paragraph{\textsc{Hamilton} vs.  energy operator}
According to classical mechanics, the \textsc{Hamilton} function of generalized coordinates $x_r$ an momenta $p_r$ of a system is equal to its entire energy:
\begin{equation}\label{Hfun}
E=\mathcal{H}(p_r,x_r)\equiv\frac{1}{2m}\sum_{r}p_r^2+U(x_r)
\end{equation}
Replacing the variables by operators and their application to a state $|\phi\ket$ leads to the relationship
\begin{equation}\label{Hop01}
\hat{E}|\phi\ket=\hat{H}|\phi\ket\equiv\left(\frac{1}{2m}\sum_{r}\hat{p}_r^2+U(\hat{x}_r)\right)|\phi\ket
\end{equation}
between the energy and the \textsc{Hamilton} operator which is nothing less than the \textsc{Schrödinger} equation in terms of matrix mechanics. Note that $\hat{H}$ and $\hat{E}$ are  \emph{essentially different} operators - \eqref{Hop01} were trivial otherwise - because $\hat{E}$ depicts the temporal behaviour of $|\phi\ket$, $\hat{H}$ its spatial behaviour and the effects of a potential. Of course, they share the same eigenfunctions $|\phi(E)\ket$ corresponding to the same eigenvalues $E$. For an $\hat{E}$- (or $\hat{H}$-)eigenstate $|\phi(E)\ket$, the operator $\hat{E}$ can be replaced by the value $E$ which leads to the stationary Schrödinger equation
\begin{equation}\label{EEWG}
\hat{H}|\phi(E)\ket=E|\phi(E)\ket.
\end{equation}
To link to wave theory, we use \eqref{ansatz2} to express the momentum operators and the energy operator in position representation:
\begin{eqnarray}
\hat{p}_r&=&~~i^{-1}\partial_{x_r}=-i\partial_{x_r}\label{pop}\\
\hat{E}&=&-i^{-1}\partial_{t}=~~i\partial_{t}\label{Eop}.
\end{eqnarray}
\paragraph{\textsc{Schrödinger} equation}
Substituting \eqref{pop} and \eqref{Eop} in \eqref{Hop01} immediately yields (\textsc{Schrödinger}, 1926)
\begin{equation}\label{schroedinger}
\hat{H}\phi(\vec{x},t)=\left(\frac{-\nabla^2}{2m}+U(\vec{x})\right)\phi(\vec{x},t)=i\frac{\partial}{\partial t}\phi(\vec{x},t).
\end{equation}
In spatial representation and using \eqref{pop}, equation \eqref{EEWG} becomes
\begin{equation}\label{pEops}
\hat{H}\phi=\left(\frac{-\nabla^{2}}{2m}+U(\vec{x})\right )\phi=E\phi
\end{equation}
whose solutions, according to \eqref{ansatz2} have the form
$
\phi(\vec{x},t)=\phi(\vec{x})\cdot e^{-iEt}
$ whose \emph{stationary} part $\phi(\vec{x})$ already solves \eqref{pEops}. Unlike the time-dependent solution which contains the factor $e^{iEt}$ but not $e^{-iEt}$, this function may be real and is to be understood as an interference of solutions of opposite momenta, i.e. a standing wave, e.g. describing a particle in a box.

\subsection{Special relativistic wave mechanics}
The quantization of SRT emanates from the \emph{relativistic energy-momentum-relationship}
(see appendix \ref{SRT}, \eqref{ArelEp}). Like in the \textsc{Schrödinger} case, replacing physical quantities by their operators leads to a differential equation (here for a free particle, \textsc{Oskar Klein, Walter Gordon}, 1926):
\begin{equation}\label{KGGhaupt}
(\hat{p}^\mu \hat{p}_\mu-m^2)\phi=(\eta^{\mu\rho}p_\mu p_\rho-m^2)\phi=0
\end{equation}
At one hand, this equation must always be satisfied. At the other, it fails to fully depict the behaviour of the most quantum systems not least for being 2nd order in all derivatives\footnote{A 2nd order equation has more solutions than a 1st order one.}. Some \emph{non-number} coefficients $\gamma^\mu$ are required to set up the following 1st order equation (\textsc{Paul Dirac}, 1928) \cite{dirac19280201, dirac19280301}:
\begin{equation}\label{Dirachaupt}
(\gamma^{\mu}\hat{p}_{\mu}-m)\phi=0.
\end{equation}
The $\gamma^\mu$ must neither be real nor complex, for squaring the operator on the left side yields
\begin{equation}\label{Diracquadrat}
\begin{aligned}
(\gamma^{\mu}\hat{p}_{\mu}-m)^2\phi&=
(\gamma^{\mu}\gamma^\rho\hat{p}_\mu\hat{p}_\rho+m^2-2m\gamma^\mu\hat{p}_\mu)\phi\\
&=(\gamma^{\mu}\gamma^\rho\hat{p}_\mu\hat{p}_\rho-m^2-2m\gamma^\mu\hat{p}_\mu+2m^2)\phi\\
&=(\gamma^{\mu}\gamma^\rho\hat{p}_\mu\hat{p}_\rho-m^2)\phi-2m\underset{=0}{\underbrace{(\gamma^\mu\hat{p}_\mu-m)\phi}}=0\\
&\Rightarrow(\gamma^{\mu}\gamma^\rho\hat{p}_\mu\hat{p}_\rho-m^2)\phi=0,
\end{aligned}
\end{equation}
where we remind the reader of the fact that $\gamma^{\mu}\gamma^\rho\hat{p}_\mu\hat{p}_\rho$ is a sum containing any pair of indices in any order. For $\phi(x^\mu)$ must also solve \eqref{KGGhaupt}, the $\gamma^\mu$  must both anti-commute pairwise to make mixed terms cancel out ans square to $\pm\hat{1}$ which generalize the numbers $\pm 1$.\footnote{For example, in an $n\times n$ matrix ring, $\hat{1}$ means the $n\times n$ unit matrix.}. Altogether, they satisfy the relationship
\begin{equation}\label{Dirackommute}
\gamma^{\mu}\gamma^{\rho}+\gamma^{\rho}\gamma^{\mu}=2\eta^{\mu\rho}\hat{1},
\end{equation}
where $\eta^{\mu\rho}$ (also see \eqref{eta}) is the metric tensor. The spatial coefficients display the same behaviour as the imaginary units of $\mathbb{H}$, the division ring of \emph{quaternions}. Indeed, the \textsc{Dirac} coefficients can be interpreted using biquaternions (i.e. quaternions over $\mathbb{C}$ instead of $\mathbb{R}$, see \ref{Biquaternionen}) in a more compact way than usually.

\section{Hypercomplex algebas and their applications to QT}
A hypercomplex algebra \emph{generalizes} (often \emph{extents}, though not always) the field $\mathbb{C}$ as an algebra and hence as a vector space over $\mathbb{R}$.
Essentially, the algebra has to be \emph{unitary} i.e. contain unity and hence $\mathbb{R}$ itself. Using a basis where 1 explicitly belongs to an element of such algebra is written as \cite{Kantor, frob19030416}
\begin{equation}
q=a_0+a_1\mathfrak{i}_1+\cdots+a_n\mathfrak{i}_n
\end{equation}
where the non-real basis elements $\mathfrak{i}_r, r=1,\dots,n$ are often called ``imaginary units'' \cite{Kantor} regardless of the rules how they are multiplied. We do not adopt this term for two reasons: The first one is that in algebra, the word ``unit'' implies the existence of a multiplicative inverse whereas an ``imaginary unit'' in the above meaning can be a \emph{zero divisor} which forbids division by them. The second has something to do with the term ``imaginary'': At least if 1 and a non-real basis element form a 2D subalgebra, this is easy to show containing a non-real element which squares to one of the real elements -1, 0 or 1; it is such an element we wish to reserve the term ``imaginary'' for.\\
However, multiplication always distributes over addition from both sides \cite{Bremner, Study} whereas any other property of multiplication like reversibility (i.e. division), associativity or even commutativity are not constitutive. These properties are exactly what the differences between algebras of the same dimension essentially consist of, for a basis transformation can alter the rules of multiplication such that it becomes at least difficult to recognize an algebra. At the other hand, different rules of multiplication don't automatically mean a different algebra.\\
In general, the product of two basis elements is a linear combination of the entire basis, i.e.
\begin{equation}\label{allgprod}
\mathfrak{i}_r\mathfrak{i}_s=\sum_{\mu=0}^{n}p_{rs\mu}\mathfrak{i}_\mu= p_{rs0}+p_{rs1}\mathfrak{i}_1+\cdots +p_{rsn}\mathfrak{i}_n
\end{equation}
where $\mathfrak{i}_0\coloneqq 1$. Note that this has nothing to do with the imaginary unit $i_0$ introduced below.
In the following, we confine our considerations to algebras which have a basis in which for any ordered pair $(r,s)$ and hence any product $\mathfrak{i}_r\mathfrak{i}_s$, there is at most one nonzero coefficient $p_{rs\mu}$, i.e.
\begin{equation}\label{spezprod}
\forall r,s\in \{1,\cdots,n\}\exists\mu\in\{0,\cdots,n\}:\mathfrak{i}_r\mathfrak{i}_s\in\{0,-\mathfrak{i}_\mu,+\mathfrak{i}_\mu\}.
\end{equation}
Of course, we are going to presume such basis as given. In this case, there are finitely many possible rules of multiplication, $(2n+3)^{n^2}$ being an upper boundary.

\paragraph{Subspaces and subalgebras}
A (proper) subspace $\mathcal{U}\subset\mathcal{A}$ is a (proper) \emph{subalgebra} of $\mathcal{A}$ iff
\begin{equation}
\forall\alpha,\beta\in\mathcal{U}: \alpha\beta\in\mathcal{U}\wedge\beta\alpha\in\mathcal{U}.
\end{equation}
\paragraph{Ideals and zero divisors}
A (proper) subalgebra $\mathcal{J}\subsetneq\mathcal{A}$ is a (proper) \emph{ideal} of $\mathcal{A}$ iff
\begin{equation}
\forall\gamma\in\mathcal{A},\beta\in\mathcal{J}: \beta\gamma\in\mathcal{J}\wedge\gamma\beta\in\mathcal{J}.
\end{equation}
An algebra is called simple iff it contains no proper ideals except of $\{0\}$.\\
Two elements $\alpha,\beta\in\mathcal{A}\setminus\{0\}$ are called \emph{zero divisors}\footnote{To speak more exactly, $\alpha$ is called a left and $\beta$ is called a right zero divisor.} iff $\alpha\cdot\beta=0$. In $\mathbb{R}$-algebras, zero divisors use to belong to ideals. It is obvious that $\alpha\in\J_1, \beta\in\J_2$ are zero divisors if $\J_1\cap\J_2=\{0\}$.
Division by $\beta\in\mathcal{J}$ is always impossible:
\begin{itemize}
\item[-]If $\gamma\notin\mathcal{J}$, the equations $\beta\xi=\gamma$ and $\xi\beta=\gamma$ have no solution $\xi\in\mathcal{A}$, namely if $\gamma=1$, i.e. there is no $\beta^{-1}$.
\item[-]If $\gamma\in\mathcal{J}$, the solution is ambiguous at least in general due to $\dim\mathcal{A}>\dim\mathcal{J}$.
\end{itemize}
We will see that zero divisors can play a vital role in \emph{eigenvalue equations} (see appendix \ref{quantSRT}, esp. \eqref{biquatspineigenvalue}).

\subsection{Familiar examples}
\subsubsection{Algebras with one imaginary unit}
Beside of $\mathbb{C}$ itself which certainly is the most famous such algebra there is also the algebra of the \emph{dual numbers} whose imaginary unit which is often called $\Omega$ squares to zero\footnote{In \cite{frob19030618}, such `numbers' are also called \textit{pseudo-nul} or roots of zero.} and the (much more interesting) algebra of the \emph{split-complex numbers} whose imaginary unit which is called $\mathcal{E}$ or $\sigma$ squares to +1; we prefer $\sigma$ due to the \textsc{Pauli} matrices which square to the $2\times 2$ unit matrix. They are also called \textit{hyperbolic numbers} due to the property
\begin{equation}
(a_0+a_1\sigma)(a_0-a_1\sigma)=a_0^2-a_1^2
\end{equation}
which is often called the \textit{modulus} and characterizes hyperbolas in the split-complex plane just like the norm of complex numbers a circle\footnote{Except split-complex numbers with modulus 0 which characterize the asymptotes of the hyperbolas and are certainly zero divisors.}. It corresponds to the square of the \textsc{Minkowski} weak norm. The algebra contains the two non-trivial (i.e. non-unity) idempotent elements
\begin{equation}
\frac{1}{2}(1\pm\sigma).
\end{equation}
These three algebras are indeed the only two-dimensional hypercomplex algebras because, for a non-real basis element $\mathfrak{i}$ with $\mathfrak{i}^2=a+b\mathfrak{i}, a,b\in\mathbb{R}$, it is easy to find an imaginary element which squares to a real number and can be normalized if non-zero \cite{Kantor}:
\begin{eqnarray}
\mathfrak{i}^2-b\mathfrak{i}+\frac{b^2}{4}=\left(\mathfrak{i}-\frac{b}{2}\right)^2&=& a+\frac{b^2}{4}\in\mathbb{R}\\
\Rightarrow\frac{\mathfrak{i}-\frac{b}{2}}{\sqrt{|a+\frac{b^2}{4}|}+\delta_{4a,-b^2}}&=&
\begin{cases}\Omega,& 4a=-b^2\\\sigma,& 4a>-b^2\\i,& 4a<-b^2\end{cases}
\end{eqnarray}

\subsubsection{Quaternions}
Unlike the examples above, the following ones contain $\mathbb{C}$ as a subalgebra and hence are really an extensions of the complex numbers.
Trying to find a reversible multiplication for 3D space vectors, \textsc{William Rowan Hamilton}, though unsuccessful in his original purpose, found the quaternions \cite{Ham01} in 1843 by adding a real component; due to him, the algebra was later called $\mathbb{H}$. There are 3 imaginary units; a quaternion $q$ is hence written as \footnote{Usually, the imaginary units are denoted by $i,j,k$ but these symbols will be used differently.}
\begin{equation}\label{quat1}
q=a_0+a_1i_1+a_2i_2+a_3i_3,\quad a_\rho\in\mathbb{R}.
\end{equation}
The rules of multiplication are summarized in Table \ref{quat}; as $\mathbb{H}$ is not commutative, the order is relevant and to be understood as row times column \cite{Ham01,Horn}.
\begin{table}[h]
\[\begin{array}{l|r|r|r|r|}
 &~~1& i_1& i_2& i_3\\\hline
1& 1& i_1& i_2& i_3\\\hline
i_1& i_1&-1& i_3&-i_2\\\hline
i_2& i_2&-i_3&-1& i_1\\\hline
i_3& i_3& i_2&-i_1&-1\\\hline
\end{array}\]
\caption{Multiplication of the quaternions}\label{quat}
\end{table}
Like in $\mathbb{C}$, every $q\in\mathbb{H}$ has a conjugate
\begin{equation}
\overline{q}=a_0-a_1i_1-a_2i_2-a_3i_3
\end{equation}
which can be used to compute
\begin{equation*}
\Re(q)=\frac{q+\overline{q}}{2},\quad \Im(q)=\frac{q-\overline{q}}{2},\quad |q|=\sqrt{q\overline{q}}.
\end{equation*}
Note the difference from $\mathbb{C}$ where, in an element $a_0+a_1i$,  it is the (real) coefficient $a_1$ which is called the imaginary part, rather than $a_1i$. For the quaternion $q$, $\Re(q), \Im(q)$ are also called its scalar and vector part.\\
A \textit{right quaternion} $q^\Im$ is defined by $\Re(q^\Im)=0$ and formally denotable as a scalar product $\vec{v}\cdot \vec{\imath}$ ($\vec{\imath}:={}^T(i_1,i_2,i_3)$). A product of two right quaternions $q_1^\Im q_2^\Im$ is
\begin{equation*}
-\vec{v}_1\cdot\vec{v}_2+\left(\vec{v}_1\times\vec{v}_2\right)\cdot \vec{\imath},
\end{equation*}
i.e. in some sense, quaternion multiplication unifies the scalar and the cross product. Quaternions can also used to describe spatial rotations\cite{Kantor}. The imaginary units share so many properties with spatial dimensions that this suggests to regard space as something \emph{essentially imaginary} - just like the imaginary \textsc{Minkowski} norms of space-like four-vectors in SRT.\\
$\mathbb{H}$ is a \emph{skew field} or \emph{division ring}, i.e. it satisfies all field axioms except of commutativity. Any plane of $\Ham$ containing $\R$ is a subalgebra isomorphic to $\C$ since the imaginary units are algebraically equivalent.
An overview of the features of $\mathbb{H}$ and other algebras is provided in appendix \ref{PropAlg}.

\subsubsection{Biquaternions}\label{Biquaternionen}
The (\textsc{Hamilton-Cayley}) biquaternions $\mathbb{C}\otimes\mathbb{H}$ are an extension of both quaternions and the bicomplex numbers discussed below. They can be perceived as an algebra over $\mathbb{C}=\Span(\{1,i_0\})$ with three `outer' imaginary units $i_1, i_2, i_3$ which anti-commute pairwise while they commute with the `inner' imaginary unit $i_0$, i.e. $i_0i_r=i_ri_0\eqqcolon \sigma_r, r=1,2,3$ for which individually
\begin{equation}
\sigma_r^2=(i_ri_0)^2=i_r^2i_0^2=(-1)\cdot(-1)=+1.
\end{equation}
Like the $i_r$, the $\sigma_r$ anti-commute pairwise which, $\varepsilon_{qrs}$ being the totally antisymmetric \textsc{Levi-Civita} pseudo-tensor, yields
\begin{equation}
\sigma_q\sigma_r=i_0^2 i_q i_r=-i_0^2 i_r i_q=-\delta_{qr}-\varepsilon_{qrs}i_s=\delta_{qr}+\varepsilon_{qrs}i_0\cdot\sigma_s.
\end{equation}
In terms of algebraic relationships, these `new' imaginary units $\sigma_r$ are isomorphic to the \textsc{Pauli} matrices and hence apt to be used in relativistic QT equations like the \textsc{Dirac} equation (see appendix \ref*{SRT}) and its non-relativistic \textsc{Pauli} approach.
An Overview of the rules of multiplication is shown in Table \ref{biquat}; as above, it is to be taken as row times column.
\begin{table}[h]
\[\begin{array}{l|r|r|r|r|r|r|r|r|}
	         &      ~~1 &      i_0 &       i_1 &       i_2 &       i_3 &  \sigma_1 &  \sigma_2 &  \sigma_3\\ \hline
	1        &        1 &      i_0 &       i_1 &       i_2 &       i_3 &  \sigma_1 &  \sigma_2 &  \sigma_3\\ \hline
	i_0      &      i_0 &       -1 &  \sigma_1 &  \sigma_2 &  \sigma_3 &      -i_1 &      -i_2 &      -i_3\\ \hline
	i_1      &      i_1 & \sigma_1 &        -1 &       i_3 &      -i_2 &      -i_0 &  \sigma_3 & -\sigma_2\\ \hline
	i_2      &      i_2 & \sigma_2 &      -i_3 &        -1 &       i_1 & -\sigma_3 &      -i_0 &  \sigma_1\\ \hline
	i_3      &      i_3 & \sigma_3 &       i_2 &      -i_1 &        -1 &  \sigma_2 & -\sigma_1 &      -i_0\\ \hline
	\sigma_1 & \sigma_1 &     -i_1 &      -i_0 &  \sigma_3 & -\sigma_2 &        ~1 &      -i_3 &       i_2\\ \hline
	\sigma_2 & \sigma_2 &     -i_2 & -\sigma_3 &      -i_0 &  \sigma_1 &       i_3 &         1 &      -i_1\\ \hline
\sigma_3 & \sigma_3 &     -i_3 &  \sigma_2 & -\sigma_1 &      -i_0 &      -i_2 &       i_1 &         1\\ \hline
\end{array}\]
\caption{Multiplication of biquaternions}\label{biquat}
\end{table}

\paragraph{Inner and outer conjugate}
For a complex number $z=x+iy, x,y\in\mathbb{R}$, its conjugate is unambiguously defined, namely by $\bar{z}=x-iy$. In principle this holds for a $q\in\mathbb{H}$ for all imaginary units are equivalent.\\
In contrast, $\mathbb{C}\otimes\mathbb{H}$ contains different types of imaginary units. Particularly, it can be understood as an algebra over $\C$ and thus a biquaternion $q=\alpha+\beta_0i_0+\sum_{r=1}^{3}(\beta_ri_r+\beta_{r+3}\sigma_r),\; \alpha,\beta_\mu\in\R$ can also be written as $q=a_0+\sum_r a_r i_r,\; a_\mu\in\C$. Beside the `plain' conjugate $\breve{q}=\alpha+\beta_0i_0+\sum_{r=1}^{3}(\beta_ri_r+\beta_{r+3}\sigma_r)$, $q$ there are hence the `outer' conjugate $\overline{q}=a_0-\sum_r a_ri_r$ and the inner conjugate $q^*=\bar{a}_0+\sum_r \bar{a}_r i_r$ as well\cite{QQM, Yongge}. Additionally, these types can be combined to $q^\dagger=\bar{a}_0-\sum_r \bar{a}_r i_r$.

\subsubsection{Bicomplex numbers}
An additional hypercomplex algebra containing $\mathbb{C}$ is the algebra $\mathbb{C}\otimes\mathbb{C}$ of the bicomplex numbers first described in 1892 by \textsc{Corrado Segre} who had studied the quaternions before. They can be regarded as complex numbers $a+i_1b\in\mathbb{C}_1, a,b\in\mathbb{C}_0:=\bra\{1,i_0\}\ket$ with the additional `inner' imaginary unit $i_0$. Unlike their superalgebra $\mathbb{C}\otimes\mathbb{H}$,  $\mathbb{C}\otimes\mathbb{C}$ is commutative \cite{Davenport, Segre} and contains only one `outer' imaginary unit which makes it principally interchangeable with the `inner' one. Altogether, the multiplication rules in canonical basis are given in Table \ref{bicomplex}.

\begin{table}[h]
\[\begin{array}{l|r|r|r|r|}
      &   ~~1&~~i_0&~i_1&~~\sigma\\\hline
     1&   ~~1&~~i_0&~i_1&~~\sigma\\\hline
   i_0&   i_0& -1&\sigma&-i_1\\\hline
   i_1&   i_1& \sigma&-1&-i_0\\\hline
\sigma&\sigma&-i_1&-i_0&~1\\\hline
\end{array}\]
\caption{Multiplication of bicomplex numbers (canonical basis)}\label{bicomplex}
\end{table}

In contrast to $\Ham$, $\C\otimes \Ham$ is not a division algebra but contain $\bra\{1,\sigma\}\ket$ as a subalgebra isomorphic to the split-complex numbers which are known to contain zero divisors. Like the latter, it contains the non-unity idempotent elements
\begin{equation}\label{idempot}
\left(\frac{1\pm\sigma}{2}\right)^2=\frac{1^2\pm 2\sigma+\sigma^2}{4}=\frac{2\pm 2\sigma}{4}=\frac{1\pm\sigma}{2},
\end{equation}
each of it belonging to a purely non-real subalgebra which is even an ideal.
An overview of $\mathbb{C}\otimes\mathbb{C}$ and other algebras is given in appendix \ref{PropAlg}.

\subsection{Hypercomplex generalizations of operations used in wave mechanics}
In the following, we are going to examine the criteria a hypercomplex-valued function must satisfy to be interpreted as a wave function in the \textsc{Schrödinger} sense:
\begin{enumerate}
\item Oscillations and waves must be expressible by exponential functions to formulate a wave function which solves the \textsc{Schrödinger} equation or/and its relativistic pendants (\textsc{Klein-Gordon}, \textsc{Dirac}).
\item A \textsc{Fourier} transform must be applicable bidirectionally to interchange between representations (e.g. $\vec{x}$, $\vec{p}$).
\end{enumerate}
To describe systems which cannot be measured directly, we \emph{additionally} demand a \emph{purely non-real} subspace (which will turn out to be a subalgebra and even an ideal) to satisfy these both conditions.
In the following, the basis elements of the demanded ideal will generally denoted by $\alpha$ and $\beta$ whose features will be examined.

\subsubsection{Algebraic conditions for wave functions and  \textsc{Schrödinger}-like equations}
\paragraph{Oscillations and series expansions}
In $\mathbb{C}$ (d.h. $\alpha=1, \beta=i$), \textsc{Euler}'s formula \[e^{ipx}=\cos(px)+i\sin(px), p,x\in\mathbb{R}\] links exponential functions to trigonometric functions and hence to oscillations which is also recognizable with help of the \textsc{Taylor} series, its even exponent summands forming the cosine series and its odd ones the sine series multiplied by~$i$:
\begin{equation}\label{Euler}
\begin{aligned}
e^{ipx}& =\sum_{n=0}^{\infty}\frac{i^n(px)^n}{n!}=\sum_{r=0}^{\infty}\left(\frac{i^{2r}(px)^{2r}}{(2r)!}+\frac{i^{(2r+1)}(px)^{(2r+1)}}{(2r+1)!}\right)\\&
=\sum_{r=0}^{\infty}(-1)^{r}\frac{(px)^{2r}}{(2r)!}+i\sum_{r=0}^{\infty}(-1)^{r}\frac{(px)^{2r+1}}{(2r+1)!}\\&=\cos (px)+ i\sin (px)
\end{aligned}
\end{equation}
In a hypercomplex algebra $\mathcal{A}$ and its subspaces/subalgebras, the series expansion can show in a corresponding manner whether an exponential function $\alpha e^{\beta px},\;\alpha,\beta\in\mathcal{A}$ describes oscillations and waves.
For this purpose, powers must be well-defined which requires $\mathcal{A}$ and its subalgebras to be at least \emph{power associative} and \emph{flexible} (see appendix \ref{assoabschwaechung}) which is automatically satisfied by \emph{alternative} and \emph{associative} algebras. We propose both power associativity and flexibility. The power series expansion $\alpha e^{\beta px}$ is
\begin{equation}\label{hkreihe}
\begin{aligned}
\alpha e^{\beta (px)}&=\alpha\sum_{n=0}^{\infty}\frac{\beta^n(px)^n}{n!}
=\alpha\sum_{r=0}^{\infty}\left(\frac{\beta^{2r}(px)^{2r}}{(2r)!}+\frac{\beta^{(2r+1)}(px)^{(2r+1)}}{(2r+1)!}\right)\\&=\alpha\sum_{r=0}^{\infty}\beta^{2r}\frac{(px)^{2r}}{(2r)!}+\alpha\beta\sum_{r=0}^{\infty}\beta^{2r}\frac{(px)^{(2r+1)}}{(2r+1)!}.
\end{aligned}
\end{equation}
To make the functions represented by \eqref{hkreihe} periodical, $\beta$ must behave like an imaginary unit in the sense of  $\mathbb{C}$, i.e., there must be $\gamma\in\mathcal{A}$ whose span is isomorphic to $\R$ and which satisfies $\beta^2=-1\cdot\gamma^2$. If so, there is also $\lambda\in\mathbb{R}$ with $\gamma^2=\lambda\gamma$. This implies  $\lambda^{-1}\gamma\eqqcolon\epsilon$ to be \emph{idempotent}, i.e. $\epsilon^m=\epsilon\forall m\in\mathbb{N}$ (including the possibility of $\epsilon=1$). Then, $\beta^2=-\lambda^2\epsilon$ and
\begin{equation}
\begin{aligned}
\alpha e^{\beta (px)}&=\alpha\epsilon\sum_{r=0}^{\infty}(-1)^r\frac{(\lambda (px))^{2r}}{(2r)!}+\frac{\alpha\beta}{\lambda}\epsilon\sum_{r=0}^{\infty}(-1)^{r}\frac{(\lambda (px))^{(2r+1)}}{(2r+1)!}\\&
=\alpha\epsilon\cos(\lambda (px))+\alpha\frac{\beta}{\lambda}\epsilon\sin(\lambda (px)).
\end{aligned}
\end{equation}
Within the first line, we used the idempotency of $\epsilon$ to factor it out thus obtaining functions of real arguments.
For simplicity, we assume $\lambda=1$. Obviously, $\Span(\{\epsilon, \beta\})$ is a subalgebra of $\mathcal{A}$ which is isomorphic to $\mathbb{C}$ and \emph{might} also contain $\alpha$ (not necessarily, as purely imaginary oscillations in $\mathbb{H}$ show).
\paragraph{The role of the idempotent element}
Idempotent elements like $\epsilon$ must be either 1 or zero divisors because
\begin{equation}\label{ipnt}
\epsilon^2=\epsilon\Rightarrow\epsilon\cdot\epsilon=1\cdot\epsilon\Rightarrow(\epsilon-1)\epsilon=0,
\end{equation}
and thus our proposal that $\Span(\{\epsilon,\beta\})$ is a purely non-real subalgebra of $\mathcal{A}$ implies that $\mathcal{A}$ cannot be a division algebra.

\paragraph{Oscillation and differential equations}
A `deeper' approach to oscillations than that via series and trigonometric functions are differential equations because they elementarily describe the behaviour of a system. A function $f(x)$ which is to depict a harmonic oscillation with $x$ being the phase must solve a differential equation of the form
\begin{equation}
\partial_x^2f(x)=-p^2f(x).
\end{equation}
If $f(x)=\alpha e^{\beta px}$ and $\alpha,\beta\in\mathcal{A}$,
\begin{equation}\label{dg1}
\partial_x^2\alpha e^{\beta px}=\alpha\beta^2p^2e^{\beta px}\overset{!}{=}-\alpha p^2e^{\beta px}\,\Rightarrow \alpha(\beta^2+1)=0,
\end{equation}
which implies $\beta^2=-1$ if $\mathcal{A}$ is simple and does not contain any zero divisors. 
\paragraph{Schrödinger equation for free particles}
The \textsc{Schrödinger} equation is a kind of wave equation which relates momentum and (in free particle case kinetic) energy. Thus, for a momentum and energy eigenstate $\phi$,
\begin{equation*}
\frac{p^2}{2m}\phi=E\phi.
\end{equation*}
Using the ansatz $\phi=\alpha e^{\beta(px-Et)}$, the first derivative with respect to $t$ is
\begin{equation}
\begin{aligned}
\partial_t\phi&=\alpha\beta(-E) e^{\beta(px-Et)}=-E\alpha\beta e^{\beta(px-Et)}\\&
=\mp E\beta\phi,\text{ if }\alpha\beta=\pm\beta\alpha,
\end{aligned}
\end{equation}
Thus $\beta^2\alpha=\alpha\beta^2=-\alpha$ leads to
\begin{equation}
\beta\partial_t\phi=\mp E\beta^2\phi=\pm E\phi,
\end{equation}
because $\phi$ contains $\alpha$ as a factor.
The 2nd derivative with respect to $x$ is
\begin{equation}
\partial_x^2\phi=\alpha\beta^2p^2e^{\beta(px-Et)}=-p^2\phi,
\end{equation}
making $\phi$ be an eigenfunction of the operator $-\partial_x^2$ corresponding to the eigenvalue $p^2$. Thus the \textsc{Schrödinger} equation takes the form
\begin{equation}\label{hkschroed}
-\frac{\partial^2}{2m\partial x^2}\phi=\pm\beta\frac{\partial}{\partial t}\phi
\end{equation}
depending on whether $\alpha$ and $\beta$ commute or anti-commute.

\paragraph{Oscillation and \textsc{Schrödinger} equation in quaternions}
The quaternions have infinitely many subalgebras which are isomorphic to $\mathbb{C}$ and hence allow oscillations; their basis elements are unity and an arbitrary \emph{unit right quaternion} which is defined by
\begin{equation*}
\vec{\imath}_a=a_1i_1+a_2i_2+a_3i_3\quad\text{with}\quad a_1^2+a_2^2+a_3^2=1.
\end{equation*}
Since the $i_r$ anti-commute pairwise, making mixed terms cancel out,
\begin{equation*}
\vec{\imath}_a^{\,2}=a_1^2i_1^2+a_2^2i_2^2+a_3^2i_3^2=(-1)a_1^2+(-1)a_2^2+(-1)a_3^2=-1,
\end{equation*}
it is isomorphic to $i\in\mathbb{C}$. A function $e^{\vec{\imath}_apx}$ thus depicts an oscillation which certainly holds for $\vec{\imath}_b e^{\vec{\imath}_a px}$ where \begin{equation*}
\vec{\imath}_b=b_1i_1+b_2i_2+b_3i_3\quad\text{mit}\quad b_1^2+b_2^2+b_3^2=1
\end{equation*}
is another unit quaternion. If, additionally, $\vec{\imath}_a\perp\vec{\imath}_b$, i.e. $\sum_{r=1}^{3}a_rb_r=0$, the oscillation takes place within a purely imaginary subspace. Such an exponential function within a purely imaginary plane may e.g. be
\begin{equation}
\begin{aligned}
i_3 e^{i_1px}&=~i_3\cdot \left(1+\frac{i_1px}{1!}-1\frac{(px)^{2}}{2!}-i_1\frac{(px)^{3}}{3!}+1\frac{(px)^{4}}{4!}+i_1\frac{(px)^{5}}{5!}+\dots\right)\\&
=~i_3+i_2\frac{px}{1!}-i_3\frac{(px)^{2}}{2!}-i_2\frac{(px)^{3}}{3!}+i_3\frac{(px)^{4}}{4!}+i_2\frac{(px)^{5}}{5!}-\dots\\&
=~i_3\cos (px)+i_2\sin (px).
\end{aligned}
\end{equation}
Of course, such a function also satisfies \eqref{dg1}. According to \eqref{hkschroed} and using $\phi=i_3e^{i_1(px-Et)}$, the pairwise anti-commutativity of the imaginary unit leads to a free particle \textsc{Schrödinger} equation
\begin{equation}
-\frac{\partial^2}{2m\partial x^2}\phi=-i_1\frac{\partial}{\partial t}\phi.
\end{equation}
Thus quaternions allow oscillations to be depicted by exponential functions and even a \textsc{Schrödinger} equation to be formulated even with a purely imaginary wave function, though with the time derivative having a negative sign in contrast to the complex case.

\subsubsection{Algebraic propositions for \textsc{Fourier} transform}
In the following, we elaborate the criteria for a \textsc{Fourier} transform to be implemented within a plane of $\mathcal{A}$ by denoting the basis elements of the plane by $\alpha$ and $\beta$ and by examinating the conditions for their multiplication rules.
\paragraph{Starting from 1D-\textsc{Fourier} transform in $\mathbb{C}$}
A function $F(x)$ can often be written as a sum of many periodic functions or at least as an integral over a continuum of functions $G(p)$:
\begin{equation}
F(x)=\frac{1}{\sqrt{2\pi}}\int G(p)e^{ipx}\,\dif p
\end{equation}
The function of amplitudes is computable via
\begin{equation}
G(p)=\frac{1}{\sqrt{2\pi}}\int F(x) e^{-ipx}\,\dif x
\end{equation}
\paragraph{Hypercomplex generalizations} In the following, the above procedure is generalized to two hypercomplex elements $\alpha$ and $\beta$ yet not specified:
\begin{equation}\label{FT-p}
F(x)=\frac{1}{\sqrt{2\pi}}\int G(p)\alpha e^{\beta px}\,\dif p
\end{equation}

\begin{equation}\label{FT-x}
G(p)=\frac{1}{\sqrt{2\pi}}\int F(x)\alpha e^{-\beta px}\,\dif x
\end{equation}
A concrete value of $F$ can be extracted via \textsc{Diracs}'s delta function\footnote{This is actually a distribution which is a specific functional acting on functions rather than on numerical values. It can be interpreted as a function via the \emph{nonstandard analysis} formulated by \textsc{Abraham Robinson} in 1961 which defines different nonzero infinitesimals and infinite elements, e.g. as a normalized \textsc{Gauß} function with an infinitesimal standard deviation.} defined by the identity
\begin{equation}
\int_{-\infty}^{\infty}f(x)\delta(x)\dif x=f(0)\quad\forall f(x)
\end{equation}
and hence
\begin{equation}
F(x)=\int_{-\infty}^{\infty}F(x^\prime)\delta (x-x^\prime)\,\dif x^\prime.
\end{equation}
Using the hypercomplexly generalized integral representation of the delta function,
\begin{equation}
\delta(x-x^\prime)=\frac{1}{\sqrt{2\pi}}\int_{-\infty}^{\infty}\alpha e^{\beta p(x-x^\prime)}\dif x^\prime,
\end{equation}
this is
\begin{eqnarray}
\int_{-\infty}^{\infty}F(x^\prime)\delta (x-x^\prime)\,\dif x^\prime
&=&\frac{1}{\sqrt{2\pi}}\int_{-\infty}^{\infty}F(x^\prime) \,\dif x^\prime\int_{-\infty}^{\infty}\alpha e^{\beta p(x-x^\prime)}\,\dif p\nonumber\\
&=&\frac{1}{\sqrt{2\pi}}\int_{-\infty}^{\infty}F(x^\prime)\alpha e^{-\beta px^\prime}\,\dif x^\prime
\int_{-\infty}^{\infty}\alpha e^{\beta px}\,\dif p\\
&=&\frac{1}{\sqrt{2\pi}}\int_{-\infty}^{\infty}G(p)\alpha e^{\beta px}\,\dif p\nonumber
\end{eqnarray}
from which following conditions for the exponential function emanate:
\begin{equation}\label{bedft}
\alpha e^{\beta p(x+x^\prime)}=\alpha e^{\beta px}\cdot \alpha e^{\beta px^\prime}
\end{equation}
\begin{eqnarray}
 e^{ix} e^{ix^\prime}&=&(\cos x+i\sin x)(\cos x^\prime+i\sin x^\prime)\nonumber\\
&=&\cos x\cos x^\prime+i\cos x\sin x^\prime+i\sin x\cos x^\prime+i\sin x\sin x^\prime\\
&=&\cos (x+x^\prime)+i\sin (x+x^\prime)\nonumber
\end{eqnarray}
\begin{eqnarray}
\alpha e^{\beta x}\alpha e^{\beta x^\prime}&=&\alpha e^{\beta (x+x^\prime)}\left[\overset{\eqref{bedft}}{=}\alpha\cos (x+x^\prime)+\beta\sin( x+x^\prime)\right]\nonumber\\
&=&(\alpha\cos x+\beta\sin x)(\alpha\cos x^\prime+\beta\sin x^\prime)\\
&=&\alpha \cos x \cos x^\prime-\alpha\sin x\sin x^\prime+\beta\sin x\cos x^\prime+\beta \sin x^\prime\cos x\nonumber\\
&=&\alpha^{2} \cos x \cos x^\prime+\beta^{2}\sin x\sin x^\prime+\beta\alpha \sin x\cos x^\prime+\beta\alpha \sin x^\prime\cos x\nonumber
\end{eqnarray}
By comparing the coefficient we obtain
\begin{equation}\label{bedft2}
\alpha^{2}=\alpha\quad\beta^{2}=-\alpha\quad\alpha\beta=\beta\alpha=\beta.
\end{equation}
Thus the subalgebra has to be isomorphic to $\mathbb{C}$ anyway, i.e. have the same rules of multiplication. For a purely imaginary subalgebra, this means that $\alpha$ must be an \emph{internal identity element} (and hence a zero divisor, due to \eqref{ipnt}).
\paragraph{Application to the quaternions}
As $\Ham$ is a \emph{division algebra}, it cannot have subalgebras with internal identity elements and so fails to satisfy our proposals for \textsc{Fourier} transforming within purely imaginary subalgebras.

\subsection{Non-real complex-isomorphic subalgebras of the bicomplex numbers}
From \eqref{idempot} we already know that the bicomplex numbers contain the non-unity  idempotent elements $\frac{1\pm\sigma}{2}$ which are both `candidates' for $k$. We choose $\frac{1+\sigma}{2}\eqqcolon k$ which makes $\frac{1-\sigma}{2}=1-k=\overline{k}$\footnote{The non-real elements $k$ and $\overline{k}=1-k$ are inter-convertible.}. Beside these elements there are $\frac{1}{2}(i_0-i_1)\eqqcolon j$ with $$\left(\frac{i_0-i_1}{2}\right)^2=\frac{i_0^2-2i_0i_1+i_1^2}{2}=\frac{-1-\sigma}{2}=-k$$ and $$\frac{1+\sigma}{2}\frac{i_0-i_1}{2}=\frac{i_0-i_1}{2}$$ and further $\frac{1}{2}(i_0+i_1)=i-j=\overline{j}$ with $$\left(\frac{i_0+i_1}{2}\right)^2=\frac{i_0^2+2i_0i_1+i_1^2}{2}=\frac{\sigma-1}{2}=k-1$$ and $$\frac{1-\sigma}{2}\frac{i_0+i_1}{2}=\frac{i_0+i_1}{2}.$$
Since $j$ and $k$ are linearly independent separately and with $1$ and $i$ as well, they can be used as basis elements instead of $i_0,i_1$. If we depict the canonical basis as orthogonal, the $j$- and $k$-axes are diagonal. Hence we refer to this basis $\{1,i,j,k\}$ shortly as a oblique basis.
These multiplication rules are listed in Table \ref{tab: multtabelle}.
\begin{table}[h]
\begin{equation*}
\begin{array}{l|r|r|r|r|}
 &~~1&~~i&~j&~~k\\\hline
1&  1&  i&~j&~k\\\hline
i&  i& -1&-k&~j\\\hline
j&  j& -k&-k&~j\\\hline
k&  k&  j&~j&~k\\\hline
\end{array}
\end{equation*}
\caption{Multiplication for the bicomplex numbers represented by its oblique basis}\label{tab: multtabelle}
\end{table}

The bicomplex numbers thus have four $\mathbb{C}$-isomorphic subalgebras, two of them being purely non-real (see Table \ref{tab: Basen}).\footnote{The angle brackets and the braces within them mean linear span and can also be denoted by $\Span(\{1,i\})$.}
\begin{table}[h]
\centering
\begin{tabular}{c|c|c}
Symbolic Denotation& canonical basis&  oblique basis\\\hline
$\mathbb{C}_1$&$\bra\{1,i_1\}\ket$&$\bra\{1,i\}\ket$\\
$\mathbb{C}_0$&$\bra\{1,i_0\}\ket$&$\bra\{1,(i-2j)\}\ket$\\
$\mathcal{J}$&$\bra\{1+\sigma,i_0-i_1\}\ket$&$\bra\{k,j\}\ket$\\
$\overline{\mathcal{J}}$&$\bra\{1-\sigma,i_0+i_1\}\ket$&$\bra\{1-k,i-j\}\ket=\bra\{\overline{k},\overline{j}\}$\\\hline
\end{tabular}
\caption{$\mathbb{C}$-isomorphic planes in $\C\otimes\C$}\label{tab: Basen}
\end{table}

The subalgebras $\mathcal{J},\overline{ \mathcal{J}}$ consist of elements which are $`\overline{\cdot}'$-conjugates of each other. Additionally, they are ideals with $\J\cap\overline{\J}=\{0\}$ which implies \[ab=0\forall a\in\mathcal{J}, b\in\overline{\mathcal{J}}\]
e.g. $1/4(1+\sigma)(1-\sigma)=k(1-k)=0$ and $1/4(i_0-i_1)(i_0+i_1)=j(i-j)=0$.\\
Below we focus on \[\Span(\{1,i\})=\mathbb{C}\quad\text{and}\quad\Span(\{1+\sigma,i_0-i_1\})=\Span(\{k,j\})=\mathcal{J}.\]

\subsection{Application of the bicomplex number to QT}
In the following, we examine how linear operators known from QM act on $\mathbb{C}$- and $\mathcal{J}$-valued wave functions. For simplicity, we focus on plane waves with a certain wave vector ($\widehat{=}$ momentum) $\vec{p}$ (wave number $p$ in 1D).
\subsubsection{Ideal-valued wave functions}
If we denote \eqref{ansatz2} by $\phi_\C$ and interpret $\mathbb{C}$ - where the values come from - as a subalgebra, its $\mathcal{J}$-valued pendant with the same $\vec{p}$ and $E$ has the form
\begin{equation}\label{ansatzk}
\phi_\J(\vec{x},t)\propto k e^{j(\vec{p}\cdot\vec{x}-Et)}.
\end{equation}
By the way, the latter equals $k e^{i(\vec{p}\cdot\vec{x}-Et)}$ as well because
\begin{equation}\label{kij}
k\cdot i=k\cdot j=j
\end{equation}
like series expansion can show. For our rather elementary consideration only requires 1D, we rewrite the functions as
\begin{eqnarray}
\phi_\C&=&e^{i(px-Et)}\\
\phi_\J&=&ke^{j(px-Et)}\overset{\eqref{kij}}{=}ke^{i(px-Et)}=k\phi_\C.
\end{eqnarray}
Both functions can be interpreted as parts of an entire wave function $\phi=\phi_\C+\phi_\J$. Even $e^{jpx}$ can be denoted by $\phi_\J$ as it is seen by the series expansion, namely $\phi_\J-k+1$.
\subsubsection{Operators and the \textsc{Schrödinger} equation}
Again we start from standard QM. The partial wave function $\phi_\C$ is the eigenfunction of the momentum operator $-i\partial_x$ corresponding to the eigenvalue $p$:
\begin{equation}
-i\partial_x\phi_\C=-i\partial_xe^{i(px-Et)}=\underset{=1}{\underbrace{-i\cdot i}}\;pe^{i(px-Et)}=p\phi_\C.
\end{equation}
The operator should also apply to the entire wave function which implies that it should apply to the partial wave function $\phi_\J$  as well; the latter turns out to be an eigenfunction of the same operator corresponding to the same eigenvalue as well:
\begin{equation}
-i\partial_x\phi_\J=-i\cdot jpke^{j(px-Et)}=-i\cdot jpe^{j(px-Et)}=pke^{j(px-Et)}=p\phi_\J=-j\partial_x\phi_\J
\end{equation}
Reversely, the $k$-fold of the momentum operator should apply to the entire wave function and therefore to $\phi_\C(x)$ as well, and via
\begin{equation}
k\cdot(-i\partial_x)\phi_\C=-j\partial_x\phi_\C=-j\cdot ipe^{i(px-Et)}=-j\cdot ipe^{i(px-Et)}=pke^{i(px-Et)}=kp\phi_\C=k\cdot-i\partial_x\phi_\C,
\end{equation}
this leads to a non-real eigenvalue $kp$. In contrast, the application of this operator to $\phi_\J$ yields
\begin{equation}
k\cdot(-i\partial_x)\phi_\J=-j\partial_x\phi_\J=-j\cdot jpke^{j(px-Et)}=-j\cdot jpe^{j(px-Et)}=pke^{j(px-Et)}=p\phi_\J=kp\phi_\J,
\end{equation}
i.e. the eigenvalue is \emph{ambiguous} inasmuch as $\phi_\J$, as an eigenfunction of the operator, can be interpreted as corresponding both to $kp$ \emph{and} to $p$.  A physical interpretation of this result will be considered in future examinations. However, the only way to obtain an \emph{unambiguously} non-real eigenvalue is to apply an $\J$-valued \emph{operator} on an at least partly $\C$-valued \emph{wave function}.\\
As an eigenfunction of the momentum operator corresponding to the eigenvalue $p$, both $\phi_\C$ and $\phi_\J$ obviously solve the \textsc{Schrödinger} equation \eqref{schroedinger}, e.g. for $U=0$:
\begin{eqnarray}
\hat{H}_{\mathbb{C}}\phi_\C&=&-\frac{\partial_x^{2}}{2m}\phi_\C=\frac{-i^2p^2}{2m}\phi_\C=\frac{p^2}{2m}\phi_\C=i\frac{\partial}{\partial t}\phi_\C=i\cdot(-i)E\phi_\C=E\phi_\C\\
\hat{H}_{\mathbb{C}}\phi_\J&=&-\frac{\partial_x^{2}}{2m}\phi_\J=\frac{-j^2p^2}{2m}\phi_\J=\frac{kp^2}{2m}\phi_\J=\frac{p^2}{2m}\phi_\J=i\frac{\partial}{\partial t}\phi_\J=i\cdot(-j)E\phi_\J=kE\phi_\J=E\phi_\J
\end{eqnarray}
Applying the $k$-fold of the \textsc{Schrödinger} equation to functions yields
\begin{eqnarray}
\hat{H}_{\mathcal{J}}\phi_\J&=&-k\frac{\partial_x^{2}}{2m}\phi_\J=-kj^2\frac{p^2}{2m}\phi_\J=k\frac{p^2}{2m}\phi_\J=j\frac{\partial}{\partial t}\phi_\J=j\cdot(-j)E\phi_\J=kE\phi_\J=E\phi_\J\\
\hat{H}_{\mathcal{J}}\phi_\C&=&-k\frac{\partial_x^{2}}{2m}\phi_\C=-ki^2\frac{p^2}{2m}\phi_\C=k\frac{p^2}{2m}\phi_\C=j\frac{\partial}{\partial t}\phi_\C=j\cdot(-i)E\phi_\C=kE\phi_\C.
\end{eqnarray}
This shows that the \textsc{Schrödinger} equation in both the $\mathbb{C}$ and the $\mathcal{J}$ form (i.e. with or without $k$ which can never be got out if once in because $\J$ is an ideal) applies to $\phi_\J$, obtaining the same ambiguity as with the momentum operators.
\paragraph{Conclusion:}
For $k\phi_\J$ and $\phi_\J$ are indistinguishable, the partial \emph{wave function} $\phi_\J$ leads to eigenvalues which can be interpreted as $k$-valued but also as real as well. It is distinguishable only whether the normal \emph{operator} or their $k$-fold are applied to the $\mathbb{C}$-valued function. For physical interpretation, this suggests to regard the operators, rather than the wave functions, as the extension of QT which is made even more plausible as far as in the description of photons\cite{Kuhn}, the carrier of the actual physical quantities like e.g. the  electric field intensity is not the wave function but the operators.
\subsubsection{Change of representation and \textsc{Fourier} transform}
Just like $\phi_\C$, $\phi_\J$ should  have a momentum and energy representation which is obtained by \textsc{Fourier} transform according to \eqref{FT-x}. In 1D which is clearly sufficient for showing it in principal, this is
\begin{equation}
\phi_\J(p)=\frac{1}{\sqrt{2\pi}}\int_{-\infty}^{\infty}\phi_\J(x)ke^{-jp^\prime x}\dif x=\frac{1}{\sqrt{2\pi}}\int_{-\infty}^{\infty}ke^{j(p-p^\prime)x-Et)}\dif x=k\delta(p-p^\prime)e^{jEt}.
\end{equation}
The delta function is obtained by, roughly\footnote{Or in terms of nonstandard analysis where infinite and infinitesimal quantities are well-defined.} speaking, phase factors cancelling out within the infinitely narrow $p$-range $p^\prime=p$, leaving the integrand constant and thus the integral infinite. This does not happen if one tries to apply the $\mathbb{C}$ \textsc{Fourier} transform to $\phi_\J$:
\begin{equation}
\frac{1}{\sqrt{2\pi}}\int_{-\infty}^{\infty}\phi_\J(x)ke^{-ip^\prime x}\dif x
=\frac{1}{\sqrt{2\pi}}\int_{-\infty}^{\infty}ke^{(jp-ip^\prime)x-jEt}\dif x
\end{equation}
Here the integrand stays periodic for $p^\prime=p$, leaving the integral bounded. This holds for applying the $\mathcal{J}$ \textsc{Fourier} transform to $\phi_\C$.
A physical interpretation of this will be given in future examinations.

\subsubsection{Double conjugate, modulus and expectation value}
In QT, the absolute square $\overline{\phi}_1(x)\phi_\C(x)$\footnote{Or $\phi_\C^*(x)\phi_\C(x)$ like usual in physics} of a wave function of an observable $x$ is interpreted as a probability density for measuring a certain value of $x$, the wave function being complex-valued and \emph{hence the conjugation unambiguously defined}. As we have seen, in $\mathbb{C}\otimes\mathbb{H}$  different types of conjugates can be defined thus and in $\mathbb{C}\otimes\mathbb{C}$ as well. Beside a ``plain'' conjugate which maps any imaginary component to its negative, there is an `outer' one which does so with $i_1$ and thus with $\sigma$ and an `inner' which maps $i_0$ and $\sigma$ to their negatives. Both map an element from $\mathcal{J}$ to one from $\overline{\mathcal{J}}$ whose product with the former is always zero, hence yielding $\overline{\phi_\J}\phi_\J\equiv 0$.\\ There is also a combined or double conjugate of $q\in\mathbb{C}\otimes\mathbb{C}$ which is defined by $q^\dagger\coloneqq \overline{q^*}=\overline{q}^*$. For $i_0^\dagger=-i_0, i_1^\dagger=-i_i$, $\sigma^\dagger=(-i_0)(-i_1)=\sigma$ and thus $k^\dagger=k, j^\dagger=-j$, the conjugate of a $\J$-valued function being like in $\mathbb{C}$ just with $k$ in the place of 1.
\\
Of course, the product $q^\dagger q$, still being non-real as it contains the factor $k$, may not be called absolute square; according to the wording for split-complex numbers, we call it the ``modulus''. The modulus of a momentum operator eigenfunction is spatially constant; its non-real value is to point up that it is not a probability density which were measurable in principal:\footnote{This does \emph{not} allow with equalize ``imaginary'' to ``not measurable'' with ``real'' to ``measurable''! The real part of a usual QT wave function is no more measurable than the imaginary part. Reversely, SRT suggests to equalize ``imaginary'' with ``space like'' when \textsc{Minkowski} norms are considered.}
\begin{equation}\label{psychfunktion}
\begin{aligned}
k e^{jpx}k e^{-jpx}&=(k\cos(px)+ j\sin(px))(k\cos(px)-j\sin(px))\\
&=~k^2\cos^{2}(px)-j^{2}\sin^{2}(px)\\
&=~k\left(\cos^{2}(px) +\sin^{2}(px))\right)\\
&=~k.
\end{aligned}
\end{equation}
The choice of the conjugate is also important for defining an \emph{expectation value} or some pendant of it, respectively. As the expectation value of the operator $-i\partial_x$ in the state $\phi_\C$ is naturally \begin{eqnarray}
\bra\phi_\C|-i\partial_x|\phi_\C\ket&= e^{-i(px-Et)}\cdot -i\partial_x e^{i(px-Et)}=e^{-i(px-Et)}\cdot -i\cdot i\cdot p\cdot e^{i(px-Et)}&= p\\
\bra\phi_\C|i\partial_t|\phi_\C\ket&= e^{-i(px-Et)}\cdot i\partial_t e^{i(px-Et)}= e^{-i(px-Et)}\cdot i\cdot-i\cdot E\cdot e^{i(px-Et)}&= E,
\end{eqnarray}
and taking $\phi^\dagger$ as the conjugate, the expectation value of the same operator in the state $\phi_\J$ is
\begin{eqnarray}
\begin{aligned}
\bra\phi_\J|-i\partial_x|\phi_\J\ket&=ke^{-j(px-Et)}\cdot -i\partial_x ke^{j(px-Et)}=kp\\
&=e^{-i(px-Et)}\cdot -j\partial_x e^{i(px-Et)}=\bra\phi_\C|-j\partial_x|\phi_\C\ket\\
&=ke^{-j(px-Et)}\cdot -j\partial_x ke^{j(px-Et)}=\bra\phi_\J|-j\partial_x|\phi_\J\ket.
\end{aligned}\\
\begin{aligned}
\bra\phi_\J|i\partial_t|\phi_\J\ket&=ke^{-j(px-Et)}\cdot i\partial_t ke^{j(px-Et)}=kE\\
&=e^{-i(px-Et)}\cdot j\partial_t e^{i(px-Et)}=\bra\phi_\C|j\partial_t|\phi_\C\ket\\
&=ke^{-j(px-Et)}\cdot j\partial_t ke^{j(px-Et)}=\bra\phi_\J|j\partial t|\phi_\J\ket
\end{aligned}
\end{eqnarray}
and thus is $k$-valued wherever the wave function or the operator is $\mathcal{J}$-valued, even where the eigenvalue is ambiguous. Note that $ke^{i(px-Et)}=ke^{j(px-Et)}$ and hence $k\phi_\C=\phi_\J$.

\section{Summary and prospects}
Initially we sketched QT in its fundamentals and saw that, in Newtonian approximation, its formulation requires complex numbers (or something isomorphic to it).\\
Further we saw that a correct and complete relativistic QT (especially the \textsc{Dirac} equation) requires even more, i.e. a higher dimensioned and non-commutative hypercomplex algebra for its coefficients.\\
Before we went into details, we first described the general properties of hypercomplex algebras. Then we considered some examples of low dimension, some of which being extensions rather than generalizations of $\mathbb{C}$. Beside the division ring or skew field  $\mathbb{H}$ of the quaternions which is by far the best known hypercomplex algebra we became acquainted with an extension of $\Ham$, namely the algebra $\mathbb{C}\otimes\mathbb{H}$ of the (\textsc{Hamilton}-\textsc{Cayley}) biquaternions which soon turned out to be apt to formulate the \textsc{Dirac} equation though some difficulties of interpretation arose which are to be concerned about in future examinations. Beside the `plain' conjugate which means to negate all imaginary components, the biquaternions provide different kinds of conjugates which we called the `inner' and the `outer' one, and their combination as well.\\
Subsequently we considered $\mathbb{C}\otimes\mathbb{C}$, a subalgebra of the bicomplex numbers which, in contrast to $\mathbb{H}$, is commutative and, like the biquaternions, contains zero divisors and hence elements division by which is impossible and some of which being idempotent that later turned out to be important.\\
Our main issue was an extension for QT with a hypercomplex algebra which at least contains one purely non-real subspace $S$ such that $S$-valued QT should be performable in the same manner as in normal complex values. This implies that $S$-valued exponential functions should describe oscillations and waves and so the formulation and solution of a \textsc{Schrödinger} equation should be able as well, which still holds for the quaternions.\\
Furthermore, it implies the possibility of \textsc{Fourier} transforms to change the basis from position to momentum representation and \textit{vice versa}. Such a purely non-real sub\emph{space} turned out to have to be a sub\emph{algebra} isomorphic to $\mathbb{C}$. From this follows the existence of an internal identity element which must be idempotent and, for being nun-unity, also a zero divisor, thus making the subalgebra $S$ be (or belong to) a proper \emph{ideal} hence denoted by $\J$. This excludes division algebras and therefore $\mathbb{H}$.\\
Last we found the bicomplex numbers to satisfy our postulates because they contain two idempotent elements $k$ and $\overline{k}$ and $j,\overline{j}$ with $j^2=-k,\overline{j}^2=-\overline{k}$ spanning the ideals $\mathcal{J}\coloneqq\Span(\{k,j\})$ and $\overline{\mathcal{J}}\coloneqq\Span(\{\overline{k},\overline{j}\})$. Additionally, $1,i,j,k$ span the entire algebra, and we use them as the new basis.\\
In the end, we introduced two partial wave functions $\phi_\C=e^{ipx}$ and $\phi_\J=ke^{jpx}$ and applied the original \textsc{Schrödinger} equation and its $\mathcal{J}$-valued version to both. The eigenvalue obtained by the application of the $\J$-valued version to $\phi_\J$ turned out to be ambiguous insofar as it can be interpreted as $k$-valued but as real-valued as well. Last we used the combined conjugate defined above to assign a nonzero modulus to $\phi_\J$ and to compute expectation values for the state $\phi_\J$ which, in contrast to the eigenvalues, all are unambiguous.\\
Future examinations will have to physically interpret the $\mathcal{J}$-valued partial wave functions and the $\mathbb{C}\otimes\mathbb{C}$-valued entire wave function according to our results here.\\
\paragraph*{Acknowledgement}
We give thanks to Hans R. Moser for the inspiring debates and some critical advice which helped us to develop this paper.

\bibliographystyle{acm}
\bibliography{Paper_ref}
\newpage
\addcontentsline{toc}{section}{APPENDIX}
\textsf{\textbf{APPENDIX}}
\appendix
\section{Superordinate properties of hypercomplex algebras}
\subsection{general properties}
Distributivity is both related to addition and multiplication \emph{at once} inasmuch as the latter distributes over the former. All other properties are related to both \emph{individually}. In the following, we consider these properties of the multiplication because in algebras, addition is always associative, commutative and reversible.
\paragraph{Distributivity}
Distributivity, which means
\begin{equation}
\begin{aligned}
a(b+c)&=ab+ac\\
(b+c)a&=ba+ca
\end{aligned}
\end{equation}
is a basic proposition for any hypercomplex algebra.
\paragraph{Associativity and its dilutions}\label{assoabschwaechung}
An algebra $\mathcal{A}$ is called associative if
\begin{equation}
(ab)c=a(bc)\quad\forall a,b,c\in\mathcal{A}.
\end{equation}
Examples are, of course, $\mathbb{R},\mathbb{C}$ and $\mathbb{H}$ and all $n\times n$ matrix rings as well. Reversely, associative hypercomplex algebras have a matrix representations \cite{Bremner}, unity being represented by $n\times n$ unit matrices. Eventual zero divisors then show up as singular matrices.
Another formulation for associativity is that the \emph{associator} $[a,b,c]\eqqcolon (ab)c-a(bc)$ vanishes.\\
$\mathcal{A}$ is called \textbf{alternative} if
\begin{equation}
(aa)b=a(ab)\quad\forall a,b\in\mathcal{A}.
\end{equation}
One example is the algebra $\mathbb{O}$ of the octonions but all associative algebras are alternative as well. The name is due to the fact that the associator is alternating, i.e. $[a,b,c]=-[a,c,b]$ and so on \cite{Bremner, Schafer}.\\
$\mathcal{A}$ is called \textbf{flexible} if
\begin{equation}
(ab)a=a(ba)\quad\forall a,b\in\mathcal{A}
\end{equation}
and \textbf{power-associative} if
\begin{equation}
a^{m+n}=(a^m)(a^n)\quad\forall a\in\mathcal{A}, m,n\in\mathbb{N}.
\end{equation}
One example is the algebra $\mathbb{S}$ of the sedenions but all alternative algebras are both flexible and power-associative as well.

\paragraph{Commutativity and anti-commutativity}
$\mathcal{A}$ is called commutative or rather anti-commutative if
\begin{equation}
ab=\pm ba\forall a,b \in\mathcal{A};
\end{equation}
this is immediately visible in the multiplication table since this is symmetric or anti-symmetric to the main diagonal. However, strict anti-commutativity does not exist in hyperkomplex algebras because they contain the real numbers which commute with any other element. Nevertheless, there will be certain anti-commuting elements unless $\mathcal{A}$ is commutative.

\paragraph{General reversibility of multiplication}
$\mathcal{A}$ is called a division algebra if
\begin{equation}
z_{1}z=z_{2}\quad\mbox{and}\quad zz_{1}=z_{2}
\end{equation}
have a unique solution $z$ for all $z_{1},z_{2}\in\mathcal{A}$. If $z_1$ is a \emph{zero divisor} and belongs to an \emph{ideal} $\mathcal{I}$, respectively, there is no solution  for $z_2\notin\mathcal{I}$ and many, often even a whole continuum of solutions for $z_2\in\mathcal{I}$.

\subsection{Properties of the algebras examined in this paper}\label{PropAlg}
For the algebras explicitly mentioned and examined in this paper, we summarize their properties, i.e. the properties of the multiplication, in a table:
\begin{table}[h]
\centering
\begin{tabular}{c|c|c|c|c|c}
name             & symbol     & distributive& associative & commutative& reversible\\\hline
complex numbers  &$\mathbb{C}$& yes & yes& yes& yes\\\hline
dual numbers     & -          & yes& yes& yes& no\\\hline
split-complex numbers   & - & yes& yes& yes& no\\\hline
quaternions     &$\mathbb{H}$& yes& yes& no& yes\\\hline
biquaternions   &$\mathbb{C}\otimes\mathbb{H}$& yes& yes& no& no\\\hline
bicomplex numbers&$\mathbb{C}\otimes\mathbb{C}$& yes& yes& yes& no\\\hline
\end{tabular}
\caption{Properties of several}
\end{table}
Biquaternions and bicomplex numbers do not form division algebras, as, particularly well seen  in the so-called oblique basis (table \ref{tab: multtabelle}) where zero divisors are basis elements denoted here as $k$ and $j$. The columns and rows for $j$ and $k$ neither contain $1$ nor $i$ but each two incidents of $j$ and $k$ instead.

\section{Formalism of QT}
\subsection{Summary of the most important basic concepts}\label{QTideas}
\paragraph{\textsc{Hilbert} spaces and quantum states}
Matrix mechanics generalizes the analytical geometry of the familiar 3D space which is a special case of vector spaces over $\R$ or $\C$ with named after \textsc{David Hilbert}: It has a scalar product and therefore the euclidean norm and additionally is \emph{complete}, i.e. all \textsc{Cauchy} sequences converge \emph{within} the space.
\footnote{
Unlike $\mathbb{Q}^3$ because there are rational \textsc{Cauchy} sequences with an irrational limit.
}
These properties are common to \emph{any} \textsc{Hilbert} space.\\
In QT, a quantum state is represented by a vector from $\mathfrak{H}$ which, according to \textsc{Paul Dirac}, is denoted by $|\phi\ket$. Any complex multiple $z|\phi\ket, z\in\mathbb{C}$ represents the same state of a particle or a system, thus the state itself
which $|\phi\ket$ represents can be identified with $\Span(|\phi\ket)\subset\mathfrak{H}$ which is actually a whole 1D subspace.\\
\textsc{Hilbert} spaces can have very different dimension including infinite and even uncountably infinite. An example for a \textsc{Hilbert} space of such dimension is the function space $L^2(\mathbb{R}^3)$ of which \emph{links to wave mechanics}: The wave function $\phi(\vec{x},t)$ is straightly a specific (here: position) representation of the quantum state $|\phi\ket$.
Position space, according to its properties, is clearly itself a \textsc{Hilbert} space but as far as this \textsc{Hilbert} space of spatial functions is concerned, it is just a kind of index set.

\subparagraph{Combination of several \textsc{Hilbert} spaces}
Tensor product $\mathfrak{H}=\mathfrak{H}^{(1)}\otimes\cdots\otimes\mathfrak{H}^{(n)}$ of $n$ \textsc{Hilbert} spaces is itself a \textsc{Hilbert} spaces,  its elements being $|\phi\ket=|\phi\ket_1\cdots|\phi\ket_n$. Note that the $\mathfrak{H}_r$ can be completely different. There are many cases where such the Combination is required to provide a complete description of particles especially if they have a \emph{spin}. For example, for a spin $\frac{1}{2}$ particle such as the electron, its spin \textsc{Hilbert} space being $\mathfrak{H}=\mathbb{C}^2$.
Thus a complete description of such a particle requires the tensor product $L^2(\mathbb{R}^3)\otimes\mathbb{C}^2$ its elements being the solutions of \textsc{Wolfgang Pauli}'s equation.

\paragraph{Dual space and scalar product}
A quantum state $|\phi\ket\in\mathfrak{H}$ corresponds to a vector $\bra\phi|$ of $\mathfrak{H}^*$, the dual space of $\mathfrak{H}$ which is actually a linear map $\mathfrak{H}\to K$, namely the map of an arbitrary vector $|\psi\ket$ to its scalar product with $|\phi\ket$ which is thus denoted by $\bra\phi|\psi\ket$. In general, $K=\mathbb{C}$. Perhaps according to duality, the complex conjugate is often denoted as $z^*$ instead of $\overline{z}$ in QT.

\subparagraph{Normalization and orthonormal basis}\label{Normierung}
As a \textsc{Hilbert} space, $\mathfrak{H}$ consists of elements which have a \emph{norm} by it can be divided to \emph{normalize} it. Hence, $L^2(\mathbb{R}^3)$ is defined by consisting of \emph{square integrable} functions $\phi(\vec{x})$  for which $\int\phi^*\phi\dif^3 x<\infty$. The function $|\phi\ket$ and $\phi(\vec{x},t)$, respectively, is called normalized if
\begin{equation}\label{norm}
\bra\phi|\phi\ket=\int_{\{\vec{x}\}}\phi^*(\vec{x},t=\mathrm{const.})\phi(\vec{x},t=\mathrm{const.})\dif^3 x=1.
\end{equation}
An orthonormal basis (ONB) or complete orthonormal system (CONS) is a basis $\{|r\ket\}$ (where $r$ belongs to an index set which may be continuous) of $\mathfrak{H}$ with
\begin{equation}
\bra r|s\ket=\delta_{rs}=\begin{cases*}
1,\;r=s\\0,\;r\neq s
\end{cases*}
\end{equation}
It is a somewhat annoying that ansatz \eqref{ansatz2} itself lacks a norm and thus does not actually belong to $L^2(\mathbb{R}^3)$. However, strictly periodical functions (i.e. such with sharply defined $\vec{p}$) are something idealized.\\
Multiplication by a, extremely flat-angle normalized function\footnote{Preferably a \textsc{Gaussian} since it is its own \textsc{Fourier} transform.} leads to a square-integrable wave function whose progress is hardly discernible from \eqref{ansatz2} over a wide range. In the following, the functions are to be assumed as normalized.

\paragraph{Operators}
The concept of a matrix is generalized in $\mathfrak{H}$ by that of a linear operator $\hat{A}$. With reference to a certain CONS $|r\ket$, $\hat{A}$ has a matrix representation $\bra r|\hat{A}|s\ket$ where $r,s$ are indices which are continuous if $\mathfrak{H}$ is a function space. If $\hat{A}$ represents an observable $A$, it is Hermitian, i.e. $\bra s|\hat{A}|r\ket=\bra r|\hat{A}|s\ket^*$ which implies $\bra r|\hat{A}|r\ket\in\mathbb{R}$; this matrix element is called the \emph{expectation value} of $\hat{A}$ in the state $|r\ket$.

\subparagraph{Eigenvalues and eigenvectors, measurements}
A quantum state $|v\ket$ for which
$\hat{A}|v\ket=a_v|v\ket$ is called an \emph{eigenstate} of $\hat{A}$ corresponding to the \emph{eigenvalue} $a_v\in A= \{a\}$ and represents a quantum state where measurements of $A$ yield the value $a_v$ without emph{principal} deviations. Of course, $a_v$ is the expectation value of $\hat{A}$ in the state $|v\ket$ as well, and $\bra v|\hat{A}|v\ket=\bra v|a_v|v\ket=a_v\bra v|v\ket=a_v$.

\paragraph{Expansion in eigenstates, \textsc{Fourier} transform}
Anything which can be measured are eigenvalues of Hermitian operators like e.g. $\hat{A}$ which holds for the case that $|\phi\ket$ not an eigenstate of $\hat{A}$ because it can be \textsc{expanded} in eigenstates of $\hat{A}$ which generalizes linear combination:
\begin{equation}\label{phiEZentw}
|\phi\ket=\sum_{a\in A} z(a)|a\ket\quad\text{bzw.}\quad|\phi\ket=\int_A z(a)|a\ket\dif a
\end{equation}
There $z(a)$ is the complex probability amplitude and $z^*(a)z(a)\equiv |z(a)|^2$ is the probability or probability density of a measurement of $a$ in the state $|\phi\ket$.
An example for the expansion of a wave function $\phi(\vec{x},t)$ in functions of the type \eqref{ansatz2} which is actually the \emph{\textsc{Fourier} transform}
\begin{equation}
\phi(\vec{p},t)=\mathcal{F}(\phi(\vec{x},t))=(2\pi)^{-\frac{3}{2}}\int_{\{\vec{x}\}}\phi(\vec{x},t)e^{-i\vec{p}\cdot\vec{x}}\dif^3x.\label{FTrp},
\end{equation}
where $\phi(\vec{p},t)$ provides the coefficients which quantify the ratio of the momentum eigenfunction for any $\vec{p}$, i.e., $|\phi(\vec{p},t)|^2$ is the probability density for a certain momentum measurement. Reversely, they can be used to re-compose the function by the inverse transform
\begin{equation}
\phi(\vec{x},t)=\mathcal{F}(\phi(\vec{p},t))=(2\pi)^{-\frac{3}{2}}\int_{\{\vec{p}\}}\phi(\vec{p},t)e^{i\vec{p}\cdot\vec{x}}\dif^3p\\\label{FTpr}.
\end{equation}

The fact that functions that are \textsc{Fourier} transforms of each other are apt to be taken as momentum and position representation of the same quantum state is due to \textsc{Marc Antoine Parseval}'s theorem which says
\begin{equation}
\int |\phi(\vec{x},t)|^2 d^3x=\int |\phi(\vec{p},t)|^2 d^3p.
\end{equation}
\paragraph{Uncertainty relation} Standard deviations of such functions are reciprocal, i.e. the \textsc{Fourier} transform of a function with a flat progression is practically zero outside of an extremely small neighbourhood of 0 but with huge values inside, being a finite approach of \textsc{Dirac}'s \emph{delta function}.
The product of these standard deviations never falls below $\hbar/2$ (in conventional units), only reaching it in the case of  \textsc{Gaussians} which are a \emph{fixed point} of the \textsc{Fourier} transform.
This relation generally applies o two observables whose operators $\hat{A}, \hat{B}$ have a fixed commutator $[\hat{A},\hat{B}]$ and hence no eigenstates in common (\textsc{Heisenberg}, 1925). If the commutator itself is an operator, there may be common eigenstates as this is the case for the components of an angular momentum, namely if $|\vec{L}|=0$.

\subsection{Examples of notation in QT}\label{anmQT}
\textsc{Dirac}'s bra-ket notation allows to denote quantum states in a very abstract and general manner which contains extreme examples like a two basis state space at one hand or a space of position wave functions containing an entire continuum of basis states at the other. We concretize the notation for both extreme cases.
\subsubsection{Two basis state system}
In this case and in matrix notation,
\begin{equation}
\bra\phi|=
\begin{pmatrix}
c_{\phi,1}^*& 
c_{\phi,2}^*
\end{pmatrix},
\quad |\psi\ket=
\begin{pmatrix}
c_{\psi,1}\\ 
c_{\psi,2}
\end{pmatrix}\quad\Rightarrow\quad
\bra\phi|\psi\ket=
\begin{pmatrix}
c_{\phi,1}^*& 
c_{\phi,2}^*
\end{pmatrix}
\begin{pmatrix}
c_{\psi,1}\\ 
c_{\psi,2}
\end{pmatrix}=
\sum_{r=1}^{n}c_{\phi,r}^*c_{\psi,r}.
\end{equation}
In such a \textsc{Hilbert} space and with respect to some given standard basis then written as \[\left\{\begin{pmatrix}1\\0\end{pmatrix},\begin{pmatrix}0\\1\end{pmatrix}\right\},\]$\hat{A}$ is a $2\times 2$ matrix $(a_{rs}),\;r,s=1,2$ and
\begin{equation}
\bra\phi|\hat{A}|\psi\ket=
\begin{pmatrix}
c_{\phi,1}^*& 
c_{\phi,2}^*
\end{pmatrix}
\begin{pmatrix}
a_{11} & a_{12} \\ 
a_{21} & a_{22}
\end{pmatrix} 
\begin{pmatrix}
c_{\psi,1}\\ 
c_{\psi,2}
\end{pmatrix}=
\sum_{r=1}^{n}\sum_{s=1}^{n}c_{\phi,r}^*a_{rs}c_{\psi,s}.
\end{equation}
If $|\phi\ket$ and $|\psi\ket$ form a basis of $\mathfrak{H}$ as well, $\hat{A}$ is represented by matrix elements
\begin{equation}
\begin{pmatrix}
\bra\phi|\hat{A}|\phi\ket & \bra\phi|\hat{A}|\psi\ket \\ 
\bra\psi|\hat{A}|\phi\ket & \bra\psi|\hat{A}|\psi\ket
\end{pmatrix},
\end{equation}
with respect to this basis, the diagonal elements being the expectation values of $\hat{A}$ in the states $\Span(|\phi\ket)$ and $\Span(|\psi\ket)$.
\paragraph{Spin system as an example}
Eigenvalues of spin direction are always projections of the spin to a given axis. The $z$ axis traditionally is the rotation axis in 3D space like it is easily seen by means of the definition of spherical coordinates. Hence it is conventional to take the orientation relatively to the $z$ axis as the standard basis. Eigenstates in other directions may be expanded in $z$ eigenstates, of course; for example, the  $y$ eigenstates are denoted by
\begin{equation}\label{yEZ}
\frac{1}{\sqrt{2}}(|+\ket\pm i|-\ket)=\frac{1}{\sqrt{2}}
\begin{pmatrix}
1\\\pm i
\end{pmatrix}
\end{equation}
according to convention. They are the eigenstate of the \textsc{Pauli} matrix $\sigma_2=\sigma_y$:
\begin{equation}
\begin{pmatrix}
0 & -i \\ 
i & 0
\end{pmatrix}\begin{pmatrix}
1\\\pm i
\end{pmatrix}
=\begin{pmatrix}
\pm 1\\ i
\end{pmatrix}.
\end{equation}
In the `+' case, the vector corresponds with the eigenvalue 1, in the `-' case with the eigenvalue -1 (the scale factor can be omitted in eigenvalue equations). These Eigenvalues are certainly the expectation values of the operator $\sigma_y$ in the eigenstates \eqref{yEZ} as well. The non-diagonal elements provide 0 because both eigenstates are orthogonal. Thus the operator is
\begin{equation}
\frac{1}{2}
\begin{pmatrix}
(\bra+|+i\bra-|)\sigma_y(|+\ket+i|-\ket) & (\bra+|+i\bra-|)\sigma_y(|+\ket-i|-\ket) \\ 
(\bra+|-i\bra-|)\sigma_y(|+\ket+i|-\ket) & (\bra+|-i\bra-|)\sigma_y(|+\ket-i|-\ket)
\end{pmatrix}
=
\begin{pmatrix}
1 & 0 \\ 
0 & -1
\end{pmatrix}
\end{equation}
with respect to the basis of its own eigenstates, exactly like 
the operator $\sigma_3$ or $\sigma_z$ in the standard basis.

\subsubsection{Position wave function}
If $\mathfrak{H}=L^2(\mathbb{R}^3(\vec{x}))$, the vectors are functions and sums become integrals:
\begin{equation}
\bra\phi|\vec{x}\ket=\phi^*(\vec{x},t),\quad\bra\vec{x}|\psi\ket=\psi(\vec{x},t)\quad\Rightarrow\quad\bra\phi|\psi\ket=\int_{\{\vec{x}\}}\phi^*(\vec{x},t)\psi(\vec{x},t)\dif^3x
\end{equation}
In this case and with respect to $|\phi\ket, |\psi\ket$, the matrix element is
\begin{equation}
\bra\phi|\hat{A}|\psi\ket=\int_{\{\vec{x}\}}\phi^*(\vec{x},t)\hat{A}\,\psi(\vec{x},t)\dif^3x
\end{equation}
or, more generally
\begin{equation}
\bra\phi|\hat{A}|\psi\ket=\int_{\{\vec{x}\}}\int_{\{\vec{x}^\prime\}}\phi^*(\vec{x},t)\bra\vec{x}|\hat{A}|\vec{x}^{\,\prime}\ket\psi(\vec{x}^{\,\prime},t)\dif^3x\dif^3x^\prime.
\end{equation}
For $|\psi\ket=|\phi\ket$, this is the expectation value. 
If $\bra\vec{x}|v\ket=\phi_v(\vec{x},t)$ is an eigenstate of $\hat{A}$ corresponding to the eigenvalue $a_v$,
\begin{equation}
\begin{aligned}
\bra v|\hat{A}|v\ket&=\int_{\{\vec{x}\}}\phi_v^*(\vec{x},t)\hat{A}\,\phi_v(\vec{x},t)\dif^3x=\int_{\{\vec{x}\}}\phi_v^*(\vec{x},t)a_v\,\phi_v(\vec{x},t)\dif^3x\\
 &=a_v\int_{\{\vec{x}\}}\phi_v^*(\vec{x},t)\phi_v(\vec{x},t)\dif^3x=a_v,
\end{aligned}
\end{equation}
exactly as it should be since the expectation value must equal the eigenvalue because it is exact in this case.

\section{Special Relativity and its quantization}\label{SRT}
\subsection{Relativity principle and Special Relativity}
One of the basic principles of classical mechanics is the \emph{relativity principle} (RP) first discovered by \textsc{Galileo Galilei}. It means that within two coordinate systems $K$ and $K^{\prime}$ relatively moving in $x$ direction the laws of mechanics are the same, or, more formally speaking, they are invariant under  \textsc{Galilei} transform which can be denoted as a matrix-vector equation
\begin{equation}\label{GTx}
\begin{pmatrix}
t^{\prime}\\
x^{\prime}
\end{pmatrix}
=
\begin{pmatrix}
~~1 & 0 \\ 
-v & 1
\end{pmatrix}
\begin{pmatrix}
t \\ 
x
\end{pmatrix},
\end{equation}
where $x$ is the only spatial dimension regarded here and $t$ and $x$ are combined to a vector which in full SR framework is called a four-vector.\\
However, \textsc{James Clerk Maxwell}'s basic equations of electrodynamics are not \textsc{Galilei} invariant and neither are the electromagnetic wave equations derived from them. This lead to the hypothesis of a \emph{luminiferous aether} which transmits light at a speed now known as $c$. This aether was thought to be at absolute rest. Within a moving frame - like earth's - the speed of light would hence vary with direction which should be measurable e.g. by interferometry. Suitable experiments, however, did not yield  any deviation from RP. To explain this, \textsc{Hendrik Antoon Lorentz} modified \eqref{GTx} step by step, finally obtaining the \textsc{Lorentz} transform
\begin{equation}\label{LTx}
\begin{pmatrix}
t^{\prime}\\x^{\prime}
\end{pmatrix}
=\gamma\begin{pmatrix}
~~1 & -\frac{v}{c^2} \\ 
-v & 1
\end{pmatrix}
\begin{pmatrix}
t\\x
\end{pmatrix}\qquad \gamma\coloneqq
\frac{1}{\sqrt{1-(\frac{v}{c})^2}}, 
\end{equation}
where $\gamma$ is called the \textsc{Lorentz} factor. Replacing $t\to ct$ makes \eqref{LTx} more symmetric, yielding
\begin{equation}\label{LTxs}
\begin{pmatrix}
ct^{\prime}\\x^{\prime}
\end{pmatrix}
=
\gamma
\begin{pmatrix}
~~1 & -\frac{v}{c} \\ 
-\frac{v}{c} & 1
\end{pmatrix}
\begin{pmatrix}
ct\\x
\end{pmatrix},\quad \text{symbolically writing}\quad \overset{\Rightarrow}{x}^\prime=\Lambda(\vec{v})\overset{\Rightarrow}{x}.
\end{equation} 

The electromagnetic wave equation and hence $c$ is invariant under \textsc{Lorentz} transform \cite{ltz1899} and so are the \textsc{Maxwell} equations. \emph{Especially, they satisfy} RP because unlike \textsc{Woldemar Voigt}'s 1887 transforms which also leave $c$ invariant, \textsc{Lorentz} transforms form a \emph{group} from which follows that the inverse of a \textsc{Lorentz} transform is also a \textsc{Lorentz} transform corresponding to the opposite velocity - symbolically speaking, $\Lambda^{-1}(\vec{v})=\Lambda(-\vec{v})$. In 1905, \textsc{Albert Einstein} based his theory of Special Relativity (SR)\cite{einst19050630} on them and also predicted the rest energy $E_0=mc^2$ which reversely provides an energy $E$ with the mass $m_E=Ec^{-2}$.\footnote{\textsc{Friedrich Hasenöhrl} had already computed a mass for cavity radiation in 1904, so the equivalence of energy and mass was new only in its  \emph{general} form.}
The universal constant $c$ is actually an artefact of the measuring system inasmuch as spatial and temporal distances are measured in different units.\footnote{Note that if horizontal distances were measured in meters whereas vertical where measured in feet, this would lead to a ``universal constant'' $\kappa=0,3048\text{~ft/m}$.}

\subsection{Covariant form and four-vectors}
In the framework of the so-called covariant formulation of SR which was later to facilitate the coordinate-independent formulation of General Relativity (GR), $ct$ or $t$ is a coordinate denoted by $x^0$ or $x_0$ which is the same for index 0. Altogether, $x^\mu={}^T(t,x,y,z)$ is called a contravariant four-vector whereas $x_\mu={^T(t,-x,-y,-z)}$ is the corresponding covariant four-vector. Both are converted into each other with help of the \emph{metric tensor}
\begin{equation}\label{eta}
\eta^{\mu\sigma}=\eta_{\mu\sigma}=\mathrm{diag}\{1,-1,-1,-1\}
\end{equation}
via $x^\mu=\eta^{\mu\rho}x_\rho$ and $x_\mu=\eta_{\mu\rho}x^\rho$, respectively. For two four-vectors $x_\mu,x_\mu^\prime$, a \textsc{Lorentz} invariant (weak) scalar product $x^\mu x_\mu^\prime=\eta^{\mu\rho}x_\mu x_\rho^\prime$ is defined. It is called weak or also improper because it lacks positive definiteness which is constitutive for proper scalar products. It induces an improper or weak norm $\|x^\mu\|=\sqrt{x^\mu x_\mu}$ first mentioned by and named after \textsc{Einstein}'s teacher \textsc{Hermann Minkowski}.\cite{mink19150615, mink19080908}

\subsection{Relativistic energy momentum relation and four-momentum}
The pendant of $x_\mu$ in momentum space is the \emph{four-momentum} $p_\mu={}^T(E,-p_x,-p_y,-p_z)$ while the pendant of $x^\mu$ is $p^\mu={}^T(E,p_x,p_y,p_z)$; the concept of the four-momentum is justified by the energy-momentum-relationship
\begin{equation}\label{ArelEp}
E^2-\vec{p}^{\,2}=p^\mu p_\mu= m^2,
\end{equation}
i.e. mass or rest energy is (or at least os proportional to) the absolute value of the four-momentum.

\subsection{Quantization of SR}\label{quantSRT}
Since a point in time $t=\mathrm{const.}$ is not well-defined in SR, normalization \eqref{norm} of a wave function for $t=\mathrm{const.}$ is replaced by a \emph{continuity equation} which is to emanate from  the basic equation like the following.
\paragraph{The \textsc{Klein} \textsc{Gordon} equation}
Even before \textsc{Schrödinger} set up the non-relativistic equation named after him, he replaced the physical quantities in \eqref{ArelEp}) by operators to set up the following differential equation (also see \eqref{KGGhaupt}):
\begin{equation}\label{KGG}
\hat{p}^\mu \hat{p}_\mu\phi=-\nabla^\mu\nabla_\mu\phi=-\square\phi:=\left(-\partial_t^2+\nabla^2\right)\phi=m^2\phi
\end{equation}
It is 2nd order in all derivations and hence there are real solutions, even time-dependent ones. These are solutions with a negative $E$ and were thus regarded as physically impossible for a long time and rejected by \textsc{Schrödinger}. However, it was examined further by \textsc{Oskar Klein} and \textsc{Walter Gordon} after whom it is now named (abbr.: KGE). Scrutiny reveals that even the solutions with a negative $E$ represent a positive energy. They are the \emph{antiparticle} solutions. \eqref{KGG} leads to the continuity equation
\begin{equation}\label{KGGcont}
\nabla^\mu\left(\phi^*\nabla_\mu\phi-\phi\nabla_\mu\phi^*\right)=\nabla^\mu\tilde{\jmath}_\mu=\partial_t\tilde{\varrho}+\nabla\cdot\tilde{\vec{\jmath}}=0
\end{equation}
which says that the four-current is a zero-divergence field. Its time component $\tilde{\jmath}_0\equiv \tilde{\varrho}$, however, is not positive definite and hence cannot be interpreted as a probability density. Of course, it neither induces preservation of particle number. Therefore $\tilde{\varrho}$ is best interpreted as a charge density or at least as a ``charge probability density''. Real solutions stand for electrically neutral \textsc{Klein Gordon} fields for which the terms in \eqref{KGGcont} vanish individually. Neutral particles completely described by KGE can hence both generated and annihilated without any violation of the equation. They are their own antiparticles like the photon. However, the latter is a quantum of a tensor field, namely of the electromagnetic one and can thus only incompletely described by KGE.

\paragraph{The \textsc{Dirac} equation}
In 1928, \textsc{Paul \textsc{Dirac}} came up with the idea of formulating a 1st order differential equation as an ansatz with initially unknown coefficients for later analysis of their required features. \cite{dirac19280201, dirac19280301} In covariant form and natural units, it is denoted by
\begin{equation}\label{Dirac}
\gamma^{\rho}\hat{p}_{\rho}\phi=m\phi.
\end{equation}
He postulated that any function $\phi$ which satisfies \eqref{Dirac} must satisfy \eqref{KGG} as well. This leads to the following commutation or rather anti-commutation relations \eqref{Dirackommute}.
From \eqref{Dirac}, the continuity equation
\begin{equation}\label{Dcont}
\nabla_\mu(\bar{\phi}\gamma^\mu\phi)=\nabla_\mu(\phi^\dagger\gamma^0\gamma^\mu\phi)=\nabla_\mu \tilde{j}^\mu=\partial_t\underbrace{(\phi^\dagger\phi)}_{\tilde{\varrho}}+\nabla\underbrace{(\phi^\dagger\vec{\alpha}\phi)}_{\tilde{\vec{\jmath}}}=0
\end{equation}
can be derived. The expression $\phi^\dagger\phi\eqqcolon\tilde{\varrho}$, the temporal component of the four-current, is positive definite and can hence be interpreted as a probability density which enables \eqref{Dcont} to express \emph{preservation of particle number}. This makes the \textsc{Dirac} equation apt to describe matter.\\
Using the biquaternionic (also see \ref{Biquaternionen}) imaginary units $\sigma_r$ usually written as complex $2\times 2$ matrices, the \textsc{Dirac} coefficients may be written more concretely as
\begin{equation}\label{dirac02}
\gamma^0=
\begin{pmatrix}
1&0\\0&-1\\
\end{pmatrix},
\gamma^r=
\begin{pmatrix}
0&\sigma_r\\-\sigma_r&0\\
\end{pmatrix}.
\end{equation}
Furthermore, some other coefficients $\gamma^0=\beta, \alpha^r=\gamma^0\gamma^r$ can be used to bring the equation into a \textsc{Schrödinger} form i.e. to solve it for the temporal derivative which facilitates the computation of the non-relativistic approach. For a particle in an electromagnetic field and using  $(\sigma_1,\sigma_3,\sigma_3)\eqqcolon\vec{\sigma}$ and the kinetic momentum $\hat{\vec{p}}-q\vec{A}\eqqcolon\hat{\vec{\pi}}$, the equation hence takes the form
\begin{equation}
\frac{i\partial}{\partial t}
\begin{pmatrix}
\phi_+\\\phi_-\\
\end{pmatrix}
=
(\beta m+\hat{1}qA_0+\vec{\alpha}\cdot\hat{1}\hat{\vec{\pi}})
\begin{pmatrix}
\phi_+\\\phi_-\\
\end{pmatrix}
=
\begin{pmatrix}
m+qA_0&\vec{\sigma}\cdot\hat{\vec{\pi}}\\
\vec{\sigma}\cdot\hat{\vec{\pi}}&-m+qA_0
\end{pmatrix}
\begin{pmatrix}
\phi_+\\\phi_-\\
\end{pmatrix}.
\end{equation}
In the limit of vanishing velocities and fields, this becomes
\begin{equation}
\frac{i\partial}{\partial t}
\begin{pmatrix}
\phi_+\\\phi_-\\
\end{pmatrix}
=
\begin{pmatrix}
m&0\\
0&-m
\end{pmatrix}
\begin{pmatrix}
\phi_+\\\phi_-\\
\end{pmatrix}.
\end{equation}
The case $E=+m$ implies $\phi_\J=0$, the case $E=-m$ implies $\phi_+=0$; hence $\phi_+$ represents matter and $\phi_-$ antimatter \cite{Alam, Rawat, vttoth2003}. In cases of high energies both occur, thus impeding a one-particle-description like in \textsc{Schrödinger} case.\\
For each case of $E=\pm m$, the \textsc{Pauli} equation can be derived which in `+' case  can be written as
\begin{equation}\label{Pauli}
i\frac{\partial}{\partial t}\xi\left(\frac{(\hat{\vec{p}}-q\vec{A})^2-q\vec{\sigma}\cdot(\nabla\times \vec{A})}{2m}+qA_0\right)\xi,\quad \xi=\phi_\C e^{-imt}
\end{equation}
The $\sigma_r$ are the components of the spin operator. If they are taken as matrices, the spin states are denoted by $\mathbb{C}^2$ vectors. If the operator components are written as biquaternions instead, the states must be biquaternions as well:
\begin{equation}\label{biquatspineigenvalue}
\sigma_r(1\pm\sigma_r)=\sigma_r\pm 1=\pm(1\pm\sigma_r)
\end{equation}
This means that $(1\pm\sigma_r)$, as a state, is an `eigen-biquaternion' of $\sigma_r$ corresponding to the eigenvalue $\pm 1$. This implies that the state biquaternion is a zero divisor, its plane, outer or inner conjugate as a `zero divisor partner'.

\paragraph{Obstacles of the interpretation}
Representing both operators and states by elements of the same algebra, i.e. the biquaternions blurs the difference between them. A further difficulty is the necessity to define scalar products and norms for zero divisors for which the multiplication with its conjugate is not useful. Furthermore, spin eigenstates of an operator for a certain direction should be able expanded in eigenstates of an operator for another. Using matrix-vector-notation, this is obtained without force whereas  biquaternions resist because the $\sigma_r$ are linearly independent.
These problems may have impeded that biquaternion formulation could have successfully competed with matrix-vector-formulation.

\section{Prefactors in differential operators in a \textsc{Hilbert} space over the ideal}
In the following, partial derivatives of functions of type \eqref{ansatz2} and \eqref{ansatzk} are provided with different pre-factors from $\mathcal{J}$ are listed (underlined results also apply for $\phi_\J\neq k\phi_\C$):
\begin{eqnarray}
+j\partial_x\phi_\C&=&+j\cdot ipe^{i(px-Et)}=-kpe^{i(px-Et)}=\underline{-kp\phi_\C}=-kp\phi_\J=-p\phi_\J\\
+j\partial_x\phi_\J&=&+j\cdot jpke^{j(px-Et)}=j^2pe^{j(px-Et)}=-kpe^{j(px-Et)}=\underline{-kp\phi_\J=-p\phi_\J}\\
-k\partial_x\phi_\C&=&-k\cdot ipe^{i(px-Et)}=-jpe^{i(px-Et)}=\underline{-jp\phi_\C}=-jkp\phi_\C=-jp\phi_\J\\
-k\partial_x\phi_\J&=&-k\cdot jpke^{j(px-Et)}=-jpe^{j(px-Et)}=-jp\phi_\C=-jkp\phi_\C=\underline{-jp\phi_\J=-ip\phi_\J}\\
+k\partial_x\phi_\C&=&+k\cdot ipe^{i(px-Et)}=jpe^{i(px-Et)}=\underline{jp\phi_\C=jkp\phi_\C}=jp\phi_\J\\
+k\partial_x\phi_\J&=&+k\cdot jpke^{j(px-Et)}=jpe^{j(px-Et)}=jp\phi_\C=jkp\phi_\C=\underline{jp\phi_\J=ip\phi_\J}.
\end{eqnarray}
The $+j$-valued operators yield a negative sign which were correct in case of the temporal derivative. The $k$-valued operators yield purely imaginary eigenvalues and therefore are anti-Hermitian regardless of their sign. If QT in $\mathcal{J}$ is to work equivalently to QT in $\mathbb{C}$, eigenvalues should be pseudo-real, i.e. $k$-valued.

\end{document}